\documentclass[a4,11pt]{article}
\usepackage[margin=2.5cm]{geometry}
\usepackage{graphicx}
\usepackage{mathtools}
\usepackage{authblk}
\usepackage{tensor}
\usepackage{esdiff}
\usepackage{tabularx}
\usepackage[dvipsnames]{xcolor}
\usepackage{subcaption}
\usepackage{amsmath, amssymb}
\usepackage[numbers,sort&compress]{natbib}

\usepackage[colorlinks=true
  ,urlcolor=blue
  ,anchorcolor=blue
  ,citecolor=blue
  ,filecolor=blue
  ,linkcolor=blue
  ,menucolor=blue
  ,pagecolor=blue
  ,linktocpage=true
]{hyperref}

\usepackage{xurl}
\usepackage{cleveref}

\newcommand{\wbd}{w_{\text{BD}}}
\newcommand{\cH}{\mathcal{H}}
\newcommand{\lb}{\bar{\lambda}}
\newcommand{\conb}{\bar{\varphi}}

\newcommand{\V}{\d^4 x \sqrt{-g}}
\newcommand{\VE}{\d^4 x \sqrt{-\tilde{g}}}
\newcommand{\p}{\partial}
\newcommand{\fb}{\bar{\varphi{}}}

\newcommand{\at}{\tilde{a}}
\renewcommand{\d}{\text{d}}
\newcommand{\te}{\tilde{t}}

\title{Bouncing and collapsing universes dual to late-time cosmological models}

\author{Dipayan Mukherjee\thanks{dipayanmkh@gmail.com}}
\author{H. K. Jassal\thanks{hkjassal@iisermohali.ac.in}}
\author{Kinjalk Lochan\thanks{kinjalk@iisermohali.ac.in}}

\affil{\emph{Indian Institute of Science Education and Research Mohali},
  \\SAS Nagar, Mohali-140306, Punjab, India}

\begin{document}
\maketitle
\begin{abstract}
  We use the Jordan frame-Einstein frame correspondence to explore dual
  universes with contrasting cosmological evolutions. We study the mapping
  between Einstein and Jordan frames where the Einstein frame universe describes
  the late-time evolution of the physical universe, which is driven by dark
  energy and non-relativistic matter. The Brans-Dicke theory of gravity is
  considered to be the dual scalar-tensor theory in the Jordan frame. We show
  that an Einstein frame universe, with cosmological evolution of the
  $\Lambda$CDM model, always corresponds to a bouncing Jordan frame universe
  governed by a Brans-Dicke theory. On the other hand, quintessence models of
  dark energy with non-relativistic matter component are shown to be always dual
  to a Brans-Dicke Jordan frame with a turn-around, i.e., a bounce or a
  collapse. The evolution of the equation of state of the quintessence field
  determines whether the turn-around is a bounce or a collapse. The point of the
  Jordan frame turn-around for all the cases can be tuned anywhere by choosing
  an appropriate Brans-Dicke parameter. This essentially leads to alternative
  descriptions of the late-time evolution of the physical universe, in terms of
  bouncing or collapsing Brans-Dicke universes in the Jordan frame. Therefore,
  the effect of dark energy can equivalently be seen as collapse of space in a
  conformally connected universe. We further study the stability of such
  conformal maps against linear perturbations. The effective bouncing and
  collapsing descriptions of the current accelerating universe may have
  interesting implications for the evolutions of perturbations and quantum
  fluctuations in the cosmological background.
\end{abstract}

\newpage
\tableofcontents

\section{Introduction}
\label{sec:expans-coll-dual-1}

It is well-known that some classes of modified theories of gravity,
such as \textit{scalar-tensor theories}~\cite{faraoni04, quiros19,
  fujii.maeda03} and $f(R)$ \textit{theories}~\cite{felice.ea10,
  sotiriou.faraoni10, nojiri.odintsov.ea17}, can be recast as Einstein
gravity with a minimally coupled scalar-field in a conformally
connected frame. The universe described by the modified gravity action
is referred to as the \textit{Jordan frame}, whereas, the universe
described by the Einstein-Hilbert action in the conformally connected
frame is called the \textit{Einstein frame}.

Einstein and Jordan frames are mathematically equivalent, they essentially
describe the same theory in terms of different dynamical
variables~\cite{faraoni99, postma.volponi14}. However, the equations of motion
in these frames may lead to drastically different evolutions of the
corresponding universes. Recent studies exploring conformally connected
universes with contrasting cosmological evolutions can be found
in~\cite{briscese.elizalde.ea07, paul.ea14, bhattacharya.ea16, wetterich13,
  wetterich14a, ijjas.ea15, fertig.ea16, Boisseau_2015, graef.ea17,
  bahamonde.odintsov.ea16, bahamonde.odintsov.ea17, francfort.ea19, bari.ea19,
  mukherjee.jassal.ea21}. In~\cite{bahamonde.odintsov.ea17}, it is demonstrated
that a decelerating Einstein frame universe can be conformally equivalent to
accelerating Jordan frame universes, governed by $f(R)$ and scalar-tensor
theories. The duality between an expanding Einstein frame and a collapsing
Jordan frame is studied in~\cite{mukherjee.jassal.ea21}. It is shown that for
some viable quintessence models in the Einstein frame, the corresponding Jordan
frame, governed by an $f(R)$ gravity theory, may have a collapsing description.
A general condition is derived to predict whether a quintessence model, with a
given time-dependent equation of state parameter, leads to such an
expansion-collapse duality between the conformally connected frames.
In~\cite{wetterich13} (also see~\cite{wetterich14a}), a cosmological model is
introduced where the Jordan frame universe is collapsing during the matter and
radiation dominated eras. The model maps to a standard quintessence field with
exponential potential in the Einstein frame. In~\cite{ijjas.ea15}, the authors
introduced ``anamorphic cosmology'', which combines features of both the
inflationary and ekpyrotic models of the early universe. The anamorphic universe
behaves like an expanding inflationary universe and a contracting ekpyrotic
universe at the same time, depending on different conformally invariant
criteria. The ``conflation'' model, introduced in~\cite{fertig.ea16}, also
combines inflationary and ekpyrotic scenarios, where the universe is contracting
and expanding in different conformal frames.

In this paper, we explore the Einstein frame-Jordan frame correspondence
focusing on the late-time cosmology, where the Jordan frame is governed by a
scalar-tensor theory action. We classify scalar-tensor theories based on whether
they lead to a collapsing Jordan frame, corresponding to an expanding Einstein
frame. The general condition for such expansion-collapse duality is then applied
to late-time cosmological models, where the Jordan frame action is considered to
be the Brans-Dicke action, and the Einstein frame is modeled to describe the
physical universe in the current era.

The late-time acceleration of the physical universe is realized in general
relativity by introducing \textit{dark energy}, an exotic fluid that violates
the strong energy condition~\cite{amendola.tsujikawa10, wang10,
  copeland.sami.ea06}. The $\Lambda$CDM ($\Lambda$ + Cold Dark Matter) model of
the current universe, often referred to as the \textit{concordance model} of
cosmology, interprets dark energy as the cosmological constant ($\Lambda$) in
the Einstein field equation \cite{padmanabhan03, carroll01, dodelson.ea21}. We
show that an Einstein frame which effectively describes the evolution of the
concordance model of cosmology, \textit{always} corresponds to a
\textit{bouncing} Jordan frame, governed by a Brans-Dicke theory. The transition
of the Einstein frame from a dust-dominated phase to an accelerating phase can
be associated with the bounce in the Jordan frame. We also show that
quintessence models of dark energy~(\cite{tsujikawa13, amendola.tsujikawa10,
  wang10}) in the Einstein frame are \emph{always} dual to Brans-Dicke Jordan
frames with a turn-around, i.e., a bounce or a collapse. Whether the Jordan
frame turn-around is a bounce or a collapse is determined by the evolution of
the equation of state of the quintessence field at the turn-around. Moreover,
the point of the Jordan frame turn-around for all these cases can be set up
\emph{anywhere} by choosing an appropriate Brans-Dicke parameter. As examples,
we demonstrate bouncing and collapsing behaviours in the Brans-Dicke Jordan
frames, corresponding to thawing and freezing types of quintessence models with
a non-relativistic matter component in the Einstein frame.

In general, bouncing models of cosmology are explored as a theory of the early
universe, alternative to the inflationary scenario~\cite{brandenberger.ea17,
  battefeld.peter15, novello.bergliaffa08, ijjas.steinhardt18}. For example,
bouncing scenarios are realized using scalar-tensor theories
in~\cite{polarski.ea22,boisseau.ea15,nojiri.odintsov.ea17}. As for any theory of
the early universe, cosmological perturbations play an important role in the
bouncing models. The statistical properties of the large-scale structure and
CMBR anisotropies as observed today, must be explained from the primordial
fluctuation near the bouncing epoch (see, for
example,~\cite{bozza06,bozza.ea05,bozza.ea05-1} and references therein). A
viable bouncing scenario in the early universe hence needs to be checked for
stability under perturbations.

In order to accommodate for cosmological perturbations in the present bouncing
model of the late-time universe, the conformal correspondence between the
physical Einstein frame and the bouncing Jordan frame should be stable in the
perturbative regime. Previous studies have pointed out that for early universe
bouncing scenarios, the linear order conformal correspondence may become
singular under certain conditions, breaking the Einstein frame-Jordan frame
duality (see, for example,~\cite{paul.ea14,bari.ea19}). We investigate the
stability of the conformal map in the present case, where the bouncing and
collapsing universes are dual to the concordance and quintessence models in the
Einstein frame. As an concrete example, we demonstrate the stability of the
conformal correspondence against linear perturbations for the concordance model.
For this case, the Jordan frame perturbations are numerically solved, first via
the Einstein frame using the conformal map, then directly in the Jordan frame.
The evolution of Jordan frame scalar perturbations obtained in these two ways
are in good agreement. This explicitly shows that the map between the linear
order perturbations in the conformally connected frames remains valid, even
through the bounce in the Jordan frame. The duality between the bouncing
universe and the $\Lambda$CDM universe is hence shown to be stable under linear
perturbations.

This paper is organized in the following way. In~\cref{sec:einst-jord-fram}, we
briefly introduce the Einstein frame-Jordan frame correspondence, where the
Jordan frame is governed by a scalar-tensor theory action. The general condition
for an expansion-collapse duality for scalar-tensor theories is obtained
in~\cref{sec:cond-expans-coll}. In~\cref{sec:conc-model:-bounce}, we show that
an Einstein frame, effectively describing the evolution of the concordance model
of cosmology, always corresponds to a bouncing Jordan frame which is governed by
a Brans-Dicke theory. We discuss the duality between Jordan frame universes with
turn-arounds and quintessence models in the Einstein frame
in~\cref{sec:quint-models:-bounc}. In~\cref{sec:de-sitter-expansion}, we briefly
discuss an example of the expansion-collapse duality where a de Sitter expansion
in the Jordan frame maps to a collapsing Einstein frame approaching the
singularity. A collapsing universe in a well-behaved gravity theory typically
supports the growth of perturbations as well as the quantum effects. We
investigate whether the duality between a collapsing and an expanding universe
survives as the perturbations evolve. At first glance, it might appear that the
role of backreactions from the perturbations in an expanding universe
diminishes. However, since its dual universe is collapsing where perturbations
are expected to grow, it is worthwhile to explore if such a map remains viable
when perturbations are included. This helps us understand if the late-time
accelerating universe has any appreciable features from the growth of
perturbations and quantum effects due to the same being true for its dual
universe. The effect of linear scalar perturbation in the conformal map is
discussed in~\cref{sec:einst-jord-frame}. We conclude with summary and
discussion in~\cref{sec:discussion}.

\textit{Throughout this paper Latin indices represent spacetime
  components, Greek indices represent spatial components. Metric
  signature is taken to be $(-,+,+,+)$.}

\section{Jordan and Einstein frames}
\label{sec:einst-jord-fram}
We begin with a brief review of Jordan and Einstein frames in the context of
scalar-tensor theories, for detailed discussion, see~\cite{faraoni04,
  fujii.maeda03, quiros19, nojiri.odintsov.ea17}.

In scalar-tensor theories, the gravity sector is governed by a scalar field
($\lambda$) along with the metric tensor field ($g_{ab}$). The action of a
general scalar-tensor theory can be written as~\cite{faraoni04,fujii.maeda03}
\begin{align}
  \label{eq:1}
  S_{J} = \int \d^4 x \sqrt{-g} \left( f(\lambda)R -
  \frac{1}{2} h(\lambda) g^{ab} \p_a \lambda \p_b \lambda - U(\lambda)
  \right),
\end{align}
where $f(\lambda)$, $h(\lambda)$, $U(\lambda)$ are arbitrary functions of the
scalar field $\lambda$. The universe described by this action is referred to as
the \textit{Jordan frame} universe. Scalar-tensor theories belong to a class of
extended theories of gravity which can be recast as general relativity, with a
canonical scalar field, in a conformally connected frame. With the following
conformal transformation~\cite{faraoni04,fujii.maeda03}
\begin{subequations}
  \label{eq:2}
  \begin{align}
    \label{eq:3}
    \tilde{g}_{ab} &= \Omega^2(x) g_{ab},\\
    \label{eq:4}
    \Omega^2 &= 16 \pi G f(\lambda),
  \end{align}
\end{subequations}
the action in~\cref{eq:1} can be written as
\begin{align}
  \label{eq:5}
  S_{J} =\int \d^4 x \sqrt{-\tilde{g}} \left[\frac{1}{16
\pi G} \tilde{R} - \frac{1}{2} K[\lambda] \tilde{g}^{ab} \p_a\lambda
\p_b\lambda - \frac {U(\lambda)}{(16 \pi G f(\lambda))^2} \right],
\end{align}
where,
\begin{align}
  \label{eq:6}
  K[\lambda] &= \frac{1}{16 \pi G f^2(\lambda)} \left(
h(\lambda)f(\lambda) + 3 f_{,\lambda}^2 \right).
\end{align}
The first term in~\cref{eq:5} is the Einstein-Hilbert action with respect to the
metric $\tilde{g}_{ab}$. The remaining terms describe a minimally coupled scalar
field, with non-canonical kinetic term. One may define a new scalar field
$\varphi$ by
\begin{align}
  \label{eq:7}
  \diff{\varphi}{\lambda} &= \sqrt{K[\lambda]},\
                            K[\lambda] > 0,
\end{align}
such that in terms of $\varphi$, the action in~\cref{eq:5}
becomes~\cite{faraoni04, fujii.maeda03}
\begin{align}
  \label{eq:8}
  S_{E} =\int \d^4 x \sqrt{-\tilde{g}} \left[\frac{1}{16
\pi G} \tilde{R} - \frac{1}{2} \tilde{g}^{ab} \p_a\varphi \p_b\varphi
- V(\varphi) \right],
\end{align}
where,
\begin{align}
  \label{eq:9}
  V(\varphi) = \frac {U(\lambda(\varphi))}{(16 \pi G
  f(\lambda(\varphi)))^2}.
\end{align}
Note that $K[\lambda]>0$ is necessary for the field $\varphi$ to be real. The
action~\eqref{eq:8} describes a minimally coupled canonical scalar field
$\varphi$, with potential $V(\varphi)$, in Einstein's gravity. The universe
governed by this action is referred to as the \textit{Einstein frame} universe.

We now move on to the expansion-collapse duality between the
conformally connected frames.

\section{Expansion-collapse duality between Einstein and Jordan frames}
\label{sec:cond-expans-coll}
As discussed above, a minimally-coupled scalar field in general relativity can
have an alternative description given by a scalar-tensor theory in the Jordan
frame. In this paper we are interested in a scenario where the Einstein frame is
the physical universe, undergoing the dark energy-dominated late time
accelerating phase, with non-negligible subdominant presence of non-relativistic
matter. Then corresponding to this Einstein frame, we seek a class of
scalar-tensor theories leading to a collapsing Jordan frame universe. Such a
class of scalar-tensor theories, if exists, can provide an effective description
of the expanding physical universe in the Einstein frame, in terms of a
collapsing one, in the Jordan frame. In this section we find a general condition
for such an expansion-collapse duality between the conformal frames.

Let us consider that both Jordan and Einstein frame spacetimes are
described by spatially flat FRW metrics, i.e.,
\begin{subequations}
  \label{eq:10}
  \begin{align}
    g_{ab} &= \text{diag} \left[-1, a^2(t),a^2(t),a^2(t)
             \right] \text{ and}\\
    \label{eq:11}
    \tilde{g}_{ab}&= \text{diag} \left[-1, \at^2(\te),
                    \at^2(\te),\at^2(\te) \right],
  \end{align}
\end{subequations}
respectively. Then, according to the conformal transformation~\eqref{eq:2},
Einstein and Jordan frame scale factors and coordinate times are related
by~\cite{faraoni04}
\begin{subequations}
  \label{eq:12}
  \begin{align}
    \label{eq:13}
    \tilde{a} &= \Omega a = \sqrt{16 \pi G f(\lambda)}
                a\\
    \label{eq:14}
    \d \tilde{t} &= \Omega \d t = \sqrt{16 \pi G
                   f(\lambda)} \d t.
  \end{align}
\end{subequations}
The Einstein frame action~\eqref{eq:8} leads to the usual Friedmann equations
\begin{subequations}
  \label{eq:15}
  \begin{align}
    \label{eq:16}
    \tilde{H}^2 &= \left( \frac{1}{\at}\diff{\at}{\te}
                  \right)^2 = \frac{\kappa^2}{3} \rho_\varphi(\at)\\
    \diff{\tilde{H}}{\te} &= - \frac{\kappa^2}{2} \left( \rho_\varphi +
                            P_\varphi \right) = - \frac{\kappa^2}{2} \rho_\varphi(1 + w_\varphi),
  \end{align}
\end{subequations}
where $\kappa^2 = 8 \pi G$, $\rho_\varphi$ and $P_\varphi$ are the energy
density and pressure associated with the scalar field $\varphi$,
\begin{subequations}
  \label{eq:17}
  \begin{align}
    \rho_\varphi &= \frac{1}{2} \left( \diff{\varphi}{\te}
                   \right)^2 + V(\varphi),\\
    P_\varphi &= \frac{1}{2} \left(
                \diff{\varphi}{\te} \right)^2 - V(\varphi),
  \end{align}
\end{subequations}
$\tilde{H}$ is the Einstein frame Hubble parameter, and
$w_\varphi=P_\varphi/\rho_\varphi$ is the equation of state parameter
corresponding to the scalar field $\varphi$. From the above relations, the time
derivative of the field can be written as
\begin{align}
  \label{eq:18}
  \diff{\varphi}{\te} = \sqrt{\rho_\varphi(1 +
  \omega_\varphi)}.
\end{align}
Starting with the relation between the scale factors in the two conformal
frames~\eqref{eq:13} and using~\cref{eq:7,eq:18}, one can find
\begin{align}
  \label{eq:19}
  \diff{a}{\at} = \frac{1}{\sqrt{16 \pi G}}
  f^{-\frac{1}{2}} \left( 1 - \frac{1}{2} \frac{f_{,\lambda}}{f}
  K^{-\frac{1}{2}}[\lambda] \sqrt{\frac{(1 +
  w_{\varphi})\rho_\varphi}{\tilde{H}^2}}\right),
\end{align}
where the subscript $(,\lambda)$ represents derivative with respect to the
Jordan frame scalar field $\lambda$. The condition for the expansion-collapse
duality between Jordan and Einstein frames is then obtained by setting
\begin{align}
  \label{eq:20}
  \diff{a}{\at} &< 0,
\end{align}
leading to
\begin{align}
  \label{eq:21}
  \frac{1}{\sqrt{16 \pi G}} f^{-\frac{1}{2}} \left( 1 -
  \frac{1}{2} \frac{f_{,\lambda}}{f} K^{-\frac{1}{2}}[\lambda]
  \sqrt{\frac{(1 + w_{\varphi})\rho_\varphi}{\tilde{H}^2}}\right) &< 0.
\end{align}
Note that $f(\lambda)>0$ is required to ensure that the conformal factor is
real, i.e. $\Omega^2>0$. Let us further consider $f_{,\lambda}>0$, then using
the Friedmann equation~\eqref{eq:16} the expansion-collapse condition can be put
in the form
\begin{align}
  \label{eq:22}
  1 + w_\varphi > \frac{2}{3} \left(\frac{fh}{f_{,\lambda}^2} + 3 \right).
\end{align}
In general, both sides of the above inequality may evolve in time. For a scalar
field in the Einstein frame, with arbitrary time-dependent $w_\varphi(\at)$, and
a scalar-tensor theory in the Jordan frame, specified by the functions
$f(\lambda), h(\lambda)$ ($f_{,\lambda} > 0$), there may exist periods of
evolution when $w_\varphi(\at)$ satisfies the above inequality. During such a
period, for an expanding (collapsing) Einstein frame universe, the corresponding
Jordan frame collapses (expands).

Here we would like to mention that in a previous
study~\cite{mukherjee.jassal.ea21}, we obtained a similar expansion-collapse
duality condition between the conformal frames, where the Jordan frame was
governed by an $f(R)$ theory. The expansion-collapse condition in the case of
$f(R)$ theories is solely determined by the Einstein frame quantities
($w_\varphi, \rho_\varphi, \tilde{H}$), the Jordan frame function $f(R)$ does
not appear explicitly in the condition. However, in the present case, one
requires the expressions for the Jordan frame functions $f(\lambda)$ and
$h(\lambda)$ in order to check the validity of the condition~\eqref{eq:22}. This
additional requirement in the case of scalar-tensor theories is expected, simply
because the Jordan frame action has three unspecified functions ($f,h, U$),
whereas, the Jordan frame action for $f(R)$ theories has a single unspecified
function, $f(R)$.

We now explore the expansion-collapse duality between an Einstein frame which
describes the late-time evolution of the physical universe, and a suitable
scalar-tensor theory in the Jordan frame. Previous studies have considered
different aspects of the duality between the standard universe and conformally
connected universes with bouncing/collapsing behaviours, governed by
scalar-tensor theories~\cite{wetterich13, wetterich14a, ijjas.ea15, fertig.ea16,
  graef.ea17}. In the present work we show that the dual universes with
turn-arounds are not exclusive to specific cosmological models, they are, in
fact, generic features of the concordance model and quintessence models of dark
energy.

\section{Concordance model: Bounce in the Jordan frame}
\label{sec:conc-model:-bounce}

The current accelerating expansion of the physical universe is considered to be
driven by a strong energy condition-violating exotic \textit{dark
  energy}~\cite{copeland.sami.ea06, amendola.tsujikawa10, wang10}. The equation
of state parameter of dark energy $w_{\text{de}}$ must be smaller than $-1/3$
and the value $w_{\text{de}} \approx -1$ is favoured by
observations~\cite{aghanim.akrami.ea20}. The cosmological constant $\Lambda$ is
the simplest implementation of dark energy within the framework of general
relativity. Although being consistent with observations, the cosmological
constant model suffers from several fine-tuning
problems~\cite{carroll01,padmanabhan03}. A wide class of dynamical models of
dark energy has been explored as an alternative for the $\Lambda$ model (see,
for example,~\cite{copeland.sami.ea06, amendola.tsujikawa10, wang10} and
references therein). A Quintessence field, i.e. a canonical scalar field
minimally coupled to gravity with a suitable equation of state parameter, is the
simplest of such dynamical models of dark energy~\cite{tsujikawa13,
  copeland.sami.ea06, chervon.yurov.ea19}. In the following discussion, we will
consider the cosmological constant model of dark energy. Quintessence models of
dark energy in this context will be discussed in the next section.

\subsection{Einstein frame: Concordance model}
\label{sec:einst-fram-conc}


We set up the Einstein frame for it to describe the current accelerating epoch
of the universe, consisting of the cosmological constant and a non-relativistic
matter component or dust, referred to as the \emph{concordance model} (see, for
example,~\cite{dodelson.ea21}). For a simple implementation of the concordance
model in the Einstein frame, we consider that the Einstein frame scalar field
\textit{effectively} describes both dark energy and non-relativistic matter in
the field equation level. This is achieved by taking the energy density of the
scalar field, $\rho_\varphi$, to be
\begin{subequations}
  \label{eq:23}
  \begin{align}
    \rho_\varphi(\tilde{a}) &= \rho_\Lambda +
                              \rho_{\text{m}} = \rho_\Lambda + \rho_{\text{m}0} \tilde{a}^{-3}\\
    \label{eq:24}
                            &= \rho_\text{c} \left( \Omega_\Lambda +
                              \Omega_{\text{m}0} \tilde{a}^{-3} \right),
  \end{align}
\end{subequations}
where $\rho_{\text{m}0}$ is the energy density of dust at the current epoch
($\at=1$), $\rho_\text{c} = 3 \tilde{H}^2_0/\kappa^2$ is the critical density of
the universe, $\tilde{H}_0 = \tilde{H}(\at=1)$,
$\Omega_\Lambda = \rho_\Lambda/\rho_\text{c}$, and
$\Omega_{\text{m}0} = \rho_{\text{m}0}/\rho_\text{c}$. With this role of the
Einstein frame scalar field, there is no need to add an extra matter component
in the Einstein frame, as the single scalar field takes into account both dark
energy, implemented by the cosmological constant, and matter. Since the scalar
field is the sole component in the Einstein frame, the Jordan frame action in
this case remains a pure gravity action, governed by only the metric $g_{ab}$
and the Jordan frame scalar field $\lambda$. The Einstein frame scalar field
$\varphi$ in this set up is hereafter referred to as the \textit{concordance}
field, in order to distinguish it from a quintessence model.

\begin{figure}
  \centering
  \includegraphics[width=.5\textwidth]{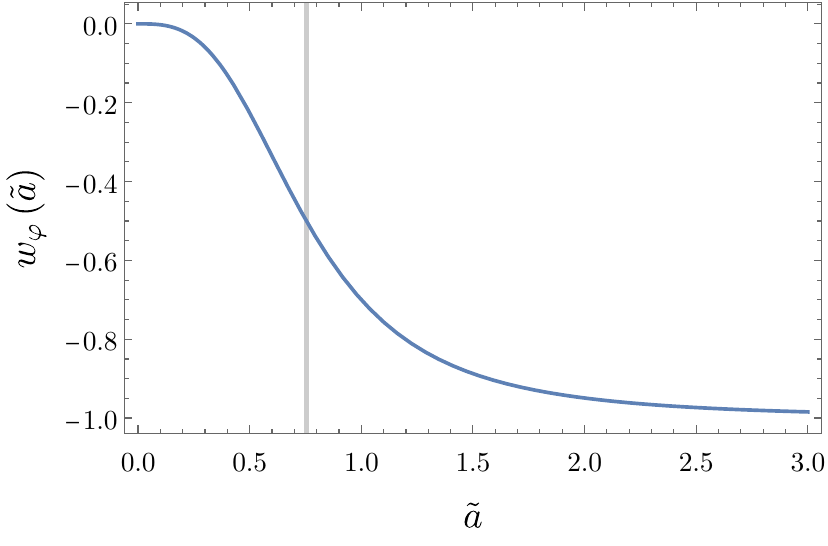}
  \caption{Equation of state parameter of the concordance field is plotted with
    respect to the Einstein frame scale factor. At the early times $\at \to 0$,
    $w_\varphi \to 0$, corresponds to the dust-dominated phase of the universe.
    In the late times $w_\varphi$ approaches $-1$, describing a
    $\Lambda$-dominated universe. The vertical line represents the period of
    dust-dark energy equivalence.}
  \label{fig:1}
\end{figure}

Starting with the above $\rho_\varphi(\at)$, one can reconstruct the action of
the concordance field as follows. Using the Friedmann equations in the Einstein
frame~\eqref{eq:15}, the equation of state $w_\varphi$, which is the effective
equation of state of the $\Lambda$ CDM model, can be written as a function of
the scale factor as (see~\cref{fig:1}),
\begin{align}
  \label{eq:26}
  w_\varphi = - 1 - \frac{2}{3}
  \frac{1}{\tilde{H}^2}\diff{\tilde{H}}{\te} = - \frac{\Omega_\Lambda
  \at^3}{1 - \Omega_\Lambda + \Omega_\Lambda \at^3}.
\end{align}
This, along with~\cref{eq:18}, leads to the concordance field as a function of
the scale factor,
\begin{align}
  \label{eq:27}
  \kappa \varphi (\tilde{a}) = \frac{1}{\sqrt{3}} \left[
  \ln \left( \sqrt{\frac{\Omega_\Lambda}{1- \Omega_\Lambda} \tilde{a}^3
  + 1} - 1 \right) -\ln \left( \sqrt{\frac{\Omega_\Lambda}{1-
  \Omega_\Lambda} \tilde{a}^3 + 1} + 1 \right) \right].
\end{align}
Finally, from~\cref{eq:17}, the concordance potential
\begin{align}
  \label{eq:28} V = \frac{1}{2}(1 - w_\varphi(\at)) \rho_\varphi(\at)
\end{align}
can be written as a function $\varphi$ as
\begin{align}
  \label{eq:29} V(\varphi) = \frac{3}{8} \frac{H_0^2}{\kappa^2}
  \Omega_\Lambda \left( 6 + \exp \left(- \sqrt{3} \kappa \varphi \right)
  + \exp\left(\sqrt{3} \kappa \varphi \right) \right).
\end{align}
where $\at$ is replaced with $\varphi$ according to~\cref{eq:27}. In other
words, a canonical scalar field with the above potential in the Einstein frame
leads to equations of motion which are \textit{exactly} the same as those of the
$\Lambda$CDM or the concordance cosmology. One can solve the Friedmann equation
for the concordance model to find the Einstein frame scale factor as
\begin{align}
  \label{eq:30}
  \at(\te) = \left( \frac{1 -
  \Omega_\Lambda}{\Omega_\Lambda}\right)^{\frac{1}{3}}
  \sinh^{\frac{2}{3}} \left( \frac{3}{2} \sqrt{\Omega_\Lambda} H_0 (\te
  - \te_i )\right),
\end{align}
where,
\begin{align}
  \label{eq:31}
  \te_i = \frac{2}{3 H_0\sqrt{\Omega_\Lambda}}
  \mathrm{arcsinh} \left( - \sqrt{ \frac{\Omega_\Lambda}{1 -
  \Omega_\Lambda}} \right),
\end{align}
which is, as expected, the scale factor of the $\Lambda$CDM model. Note that the
origin of the Einstein frame coordinate time is chosen to be the current epoch
$\te_0$, i.e.\ $\te_0=0$ and $\at(\te=\te_0=0)=1$, whereas, at $\te=\te_i$ the
scale factor $\at(\te=\te_i) = 0$.

Having set up the Einstein frame, we will now move on to a suitable Jordan frame
description of the universe.

\subsection{Jordan frame: Brans-Dicke theory}
\label{sec:jordan-frame:-brans}
The concordance scalar field in the Einstein frame, with
potential~\eqref{eq:29}, can be mapped to a wide class of scalar-tensor
theories, depending on choices of the functions $f(\lambda),\ h(\lambda)$ in the
Jordan frame action~\eqref{eq:1}. This correspondence is given by the
relations~\eqref{eq:7} and~\eqref{eq:9}, as discussed before. We are interested
in an example of the scalar-tensor theories, dual to the concordance model, such
that the conformally connected frames possess the expansion-collapse duality
feature as discussed in~\cref{sec:cond-expans-coll}.

The widely-studied Brans-Dicke theory of gravity is the prototype of
scalar-tensor theories~\cite{brans.dicke61, avilez.skordiz.ea14,li.wu.ea13,
  bonino.ea20,faraoni04,fujii.maeda03, kazempour.ea22}. A Brans-Dicke theory
dual to the concordance model may lead to the expansion-collapse duality under
certain conditions, as we will see. Apart from this, the Brans-Dicke action is
simple enough for the equations of motion in the Jordan frame to be solved
analytically. For this reason, we will consider the Brans-Dicke theory as an
example of scalar-tensor theories in the Jordan frame for the rest of the
discussion.

With the following choice of the functions~\cite{faraoni04,
fujii.maeda03},
\begin{subequations}
  \label{eq:32}
  \begin{align}
    f(\lambda) &= \frac{\lambda}{16 \pi},\\
    \label{eq:33}
    h(\lambda) &= \frac{w_{\text{BD}}}{8 \pi \lambda},
  \end{align}
\end{subequations}
the Jordan frame action~\eqref{eq:1} becomes the Brans-Dicke action
\begin{align}
  \label{eq:34}
  S_{J}^{\text{BD}} = \int \d^4 x \sqrt{-g} \left(
  \frac{\lambda}{16 \pi} R - \frac{w_{\text{BD}}}{16 \pi \lambda} g^{ab}
  \p_a \lambda \p_b \lambda - U(\lambda) \right),
\end{align}
where $\wbd$, the constant Brans-Dicke parameter, must satisfy $\wbd> - 3/2$ in
order for the concordance field $\varphi$ to be real (see~\cref{eq:7}). Note
that $f_{,\lambda} = \frac{1}{16 \pi}$, $f_{,\lambda}>0$ is ensured for all
$\lambda$. One can then apply~\cref{eq:22} to determine the condition for a
collapsing Jordan frame,
\begin{align}
  \label{eq:35}
  w_\varphi(\at) &> \frac{4}{3} w_{\text{BD}} + 1.
\end{align}
\begin{figure}
  \centering
  \includegraphics[width=.5\textwidth]{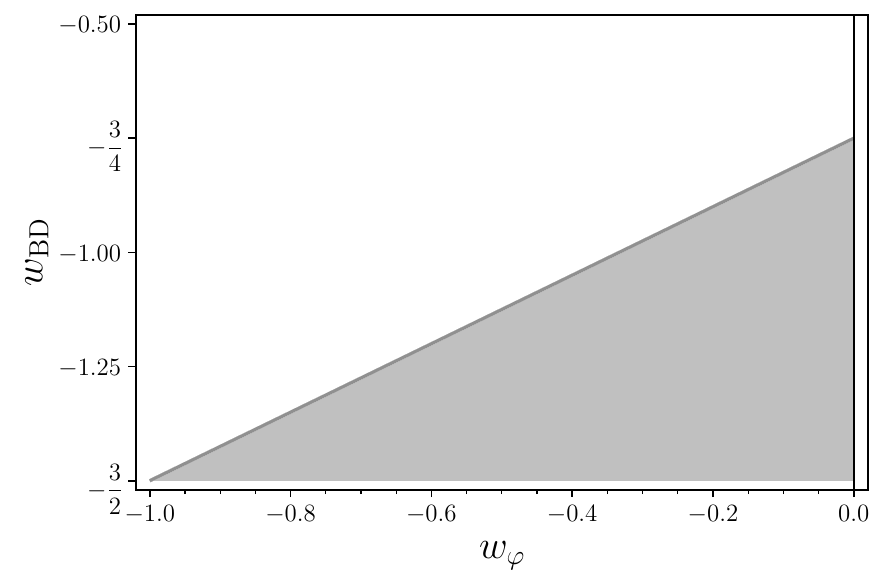}
  \caption{Condition for a collapsing Brans-Dicke Jordan frame, dual to the
    concordance model, determined by $\wbd$ and $w_\varphi$; where $\wbd$ is the
    constant Brans-Dicke parameter specifying the Jordan frame, $w_\varphi$ is
    the dynamic equation of state of the concordance field. At a given instant,
    if a pair $(w_\varphi, \wbd)$ lies in the shaded region, then the
    corresponding Brans-Dicke universe is collapsing at that instant.}
  \label{fig:2}
\end{figure}
For the Brans-Dicke model, the expansion-collapse duality condition becomes
significantly simple, as it only requires the knowledge of the equation of state
$w_\varphi(\at)$. Given the evolution of $w_\varphi(\at)$, the expanding and
collapsing phases of the Jordan frame, determined by the
condition~\eqref{eq:35}, can be visualized in~\cref{fig:2}. The shaded region in
the figure depicts the domain in the $(w_\varphi, \wbd)$ space, where the
condition~\eqref{eq:35} is satisfied. That is, for a Brans-Dicke Jordan frame,
specified by $\wbd$, at a given value of the Einstein frame scale factor
$\at_*$, if the pair $(w_\varphi(\at_*), \wbd)$ lies within the shaded region,
then we can conclude that the Jordan frame is collapsing at $\at = \at_*$.

The Einstein frame is matter dominated at the early-times; i.e.~for $\at \to 0$,
the equation of state parameter of the concordance field $w_\varphi \approx 0 $
(see~\cref{fig:1}). Eventually, as the scale factor increases, $w_\varphi$ decreases.
In the late-time of the Einstein frame, i.e., as $\at \to \infty$,
$w_\varphi \to -1$ depicting the dark-energy dominated era. We see from~\cref{fig:2}
that for $\wbd > -3/4$, the Jordan frame is never collapsing. For
$-3/2 < w_{\text{BD}} <-3/4$, the points with $w_\varphi = 0$ always lie in the shaded
region. This implies that at the beginning of the matter-dominated era of the
Einstein frame ($w_\varphi \approx 0$), the Jordan frame is collapsing. For the same value
of $\wbd$, $w_\varphi$ will then monotonically decrease towards $-1$ with increasing
$\at$. Thus, at some time, the trajectory of the Einstein frame universe in the
$(w_\varphi, \wbd)$ space will inevitably come out of the shaded region and enter the
white region, where the Jordan frame is expanding. The transition of the
trajectory from the shaded region (collapsing Jordan frame) to the white region
(expanding Jordan frame), represents a \emph{bounce in the Jordan frame}. The
Einstein frame scale factor at the point of the bounce, $\at_b$, satisfies
\begin{align}
  \label{eq:36}
  w_\varphi(\at_b) = \frac{4}{3} w_{\text{BD}} + 1.
\end{align}

\begin{figure}
  \centering
  \includegraphics[width=.6\textwidth]{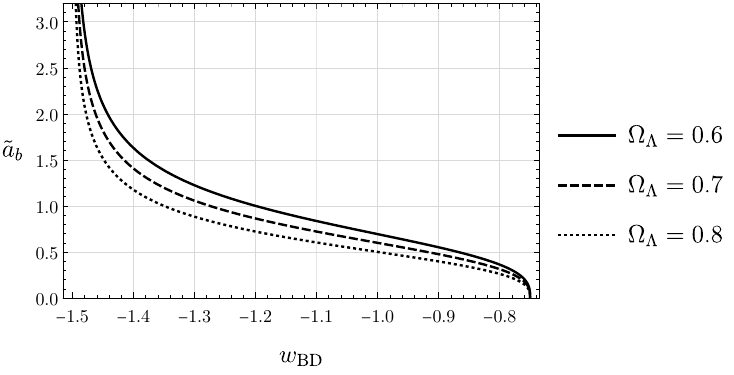}
  \caption{Einstein frame scale factor at the point of the Jordan frame bounce
    $\at_b$ is plotted with respect to the Brans-Dicke parameter $\wbd$.
    Different plots are for $\Omega_\Lambda = 0.6, 0.7, 0.8$. $\at_b$ can take any possible
    value depending on $- \frac{3}{2} < \wbd < - \frac{3}{4}$. For smaller
    values of $\wbd$, the Jordan frame bounce occurs at later times in the
    Einstein frame.}
  \label{fig:4}
\end{figure}

For the concordance scalar field in the Einstein frame, the equations of motion
of the corresponding Jordan frame can be solved analytically. Starting
with~\cref{eq:7} and using~\cref{eq:32}, one can write the Brans-Dicke field
$\lambda$ as a function of the concordance field as,
\begin{align}
  \label{eq:37}
  \lambda(\varphi) = \lambda_0 \exp \left(
  \sqrt{\frac{2}{\varpi}} \kappa \varphi \right),
\end{align}
where we define
\begin{align}
  \label{eq:38}
  \varpi = 2 \wbd +3,
\end{align}
and $\lambda_0$ is an integration constant such that
$\lambda(\varphi=0)=\lambda_0$. Given the concordance potential~\eqref{eq:29}, one can obtain the
corresponding Brans-Dicke potential using~\cref{eq:37,eq:9} as
\begin{align}
  \label{eq:39}
  U(\lambda) = \frac{3 G}{64 \pi} H_0^2 \Omega_\Lambda
  \lambda^2 \left( 6 + \left( \frac{\lambda}{\lambda_0}
  \right)^{\sqrt{\frac{3}{2}\varpi}} + \left( \frac{\lambda}{\lambda_0}
  \right)^{-\sqrt{\frac{3}{2}\varpi}} \right).
\end{align}
Now, from~\cref{eq:13}, the Brans-Dicke frame scale factor $a$ is related to the
Einstein frame scale factor as
\begin{align}
  \label{eq:40}
  a = \frac{\at} {\sqrt{G \lambda }}.
\end{align}
Using~\cref{eq:37,eq:27}, we replace $\lambda$ with $\at$ in the above expression to
write the Jordan frame scale factor as a function of the Einstein frame scale
factor,
\begin{align}
  \label{eq:41}
  a(\at) = \sqrt{\frac{8 \pi }{\kappa^2 \lambda_0}}
  \tilde{a} \left(\frac{\sqrt{\frac{ \Omega _{\Lambda }}{1-\Omega
  _{\Lambda }}\tilde{a}^3 + 1} + 1}{\sqrt{\frac{ \Omega _{\Lambda
  }}{1-\Omega _{\Lambda }}\tilde{a}^3+1}-1}\right)^{\frac{1}{\sqrt{ 6
  \varpi}}}.
\end{align}
\begin{figure}
  \centering
  \includegraphics[width=.6\textwidth]{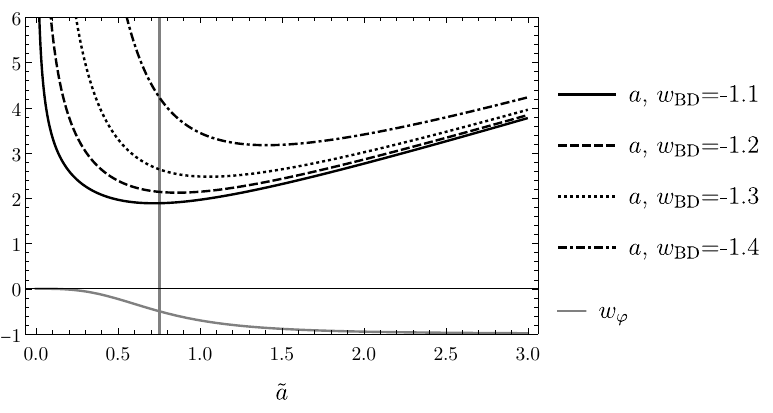}
  \caption{Bouncing behaviour of the Brans-Dicke Jordan frame, dual to the
    concordance model. Scale factors of different Jordan frames, specified by
    different $\wbd$ values, are plotted with respect to the Einstein frame
    scale factor. The gray plot below is the equation of state of the
    concordance field $w_\varphi$ , The Vertical gray line represents the epoch of
    dust-$\Lambda$ equivalence in the Einstein frame. The occurrence of the Jordan
    frame bounce shifts towards future with decreasing $\wbd$.}
  \label{fig:3}
\end{figure}
Evolution of the Jordan frame scale factor $a(\at)$ for different values of
$\wbd$ is shown in~\cref{fig:3}. For all the plots, the Jordan frame universe is
seen to be collapsing at the early matter-dominated phase in the Einstein frame.
Eventually, the Jordan frame goes through a non-singular bounce and starts
expanding with the Einstein frame. One can find the Einstein frame scale factor
at the point of bounce in the Jordan frame to be
\begin{align}
  \label{eq:42}
  \tilde{a}_{b} = \left( \frac{1 - \Omega_\Lambda}{ 2
  \Omega_\Lambda} \right)^{\frac{1}{3}} \left( - \frac{3 + 4 \wbd}{3 + 2
  \wbd} \right)^{\frac{1}{3}},
\end{align}
for $- \frac{3}{2} < \wbd < - \frac{3}{4}$ (see~\cref{fig:2}). We see, depending
on $\wbd$, the Jordan frame bounce can occur corresponding to a value of the
Einstein frame scale factor anywhere within $\at_b \to 0$ (for
$\wbd \to - \frac{3}{4}$) and $\at_b \to \infty$ (for $\wbd \to - \frac{3}{2}$),
i.e., anywhere within the entire concordance model era. The time of the Jordan
frame bounce shifts to the future with increasing value of $\wbd$ parameter, as
it can be seen from~\cref{fig:4}. For a given Brans-Dicke model ($\wbd$), the
size of the Jordan frame universe at the point of the bounce depends on the dark
energy content in the Einstein frame universe ($\Omega_\Lambda$). Interestingly,
even though the concordance model possesses a big bang-like cosmological
singularity, i.e., the Einstein frame scale factor $\tilde{a} \to 0$ at a finite
coordinate time (\cref{eq:31}), there is no such singularity in the dual
bouncing description, as it can be seen from~\cref{fig:3}. Therefore, the
appearance of the cosmological singularity can be attributed to the choice of
the conformal frame. The conformal frame-dependence of cosmological
singularities has been argued in the literature previously, for example
in~\cite{wetterich13}, the present result agrees with such a conjecture.

Note that, a similar result was obtained in~\cite{Boisseau_2015}, where, the
conformal duality was established between a bouncing Jordan frame and an
Einstein frame governed by a scalar field with a quartic potential and a
cosmological constant. In the following section, we generalize this result by
considering the case of generic quintessence models with arbitrary potentials.

It is interesting to note that depending on the choice of the constant
$\lambda_0$ in~\cref{eq:41}, the Jordan frame scale factor can become arbitrarily
small at the bounce. However, the bounce can be arranged near the current epoch
in the Einstein frame, when the Einstein fame scale factor remains $\at \sim 1$. In
this scenario, if the Jordan frame scale factor decreases below a sufficiently
small scale near the bounce, one may speculate the quantum effects in the Jordan
frame to become non-trivial; for example, quantum fluctuations in relevant
Jordan frame cosmological operators may grow in this regime. At the same time,
the quantum effects in the corresponding Einstein frame is expected to be
suppressed due to its relatively large scale. In this case, the conformal
correspondence seems to provide a map between a quantum-corrected universe and a
universe with negligible quantum effects. Alternatively, one may find the
classical conformal map to break down near the bouncing phase~\cite{pandey.ea17,
  banerjee2016, mukherjee2023}. For example,~\cite{mukherjee2023} shows that in
the absence of additional matter components, the conformal correspondence
between an expanding and collapsing universe survives at the quantum level.
Interestingly, quantum fluctuations in different cosmological operators increase
both in the collapsing and the expanding frame in similar ways, regardless of
the cosmological evolutions therein.

\section{Quintessence models: Turn-around in the Jordan frame}
\label{sec:quint-models:-bounc}

In the previous section, we demonstrated the duality between the physical
universe and bouncing universes by considering the example of the concordance
scalar field in the Einstein frame, i.e., a single canonical scalar field that
reproduces the background evolution of the $\Lambda$CDM cosmology. We now extend the
analysis to general quintessence models with other matter components, such as
non-relativistic matter and radiation, added separately. As before, the Einstein
frame represents the physical universe, therefore we wish these matter
components to be minimally coupled in the Einstein frame. This can be achieved
by considering the Jordan frame action with the following form
\begin{align}
  \label{eq:43}
  S_{J}^{BD} &= \int \d^4 x \sqrt{-g} \left( \frac{\lambda}{16 \pi} R - \frac{w_{\text{BD}}}{16 \pi \lambda} g^{ab} \p_a \lambda \p_b \lambda - U(\lambda) \right) \\
           &+ \int \V \Omega^4(\lambda) \mathcal{L}_M \left(\psi_M;\Omega^2(\lambda)g_{ab}\right),\nonumber
\end{align}
where the first term is the Brans-Dicke action same as before; in the second
term we introduce the action of a non-minimally coupled matter field $\psi_M$. The
conformal transformation $\tilde{g}_{ab} = \Omega^2 g_{ab}$ removes the
$\lambda$ dependency from the matter action, leading to the Einstein frame action
\begin{align}
  \label{eq:44}
  S_E &= \int \VE \frac{\tilde{R}}{2 \kappa^2} - \int \VE \left( \frac{1}{2} \tilde{g}^{ab} \p_a \varphi \p_b \varphi  + V(\varphi) \right) \\
      &+ \int \VE \mathcal{L}_M \left(\psi_M; \tilde{g}_{ab}   \right) . \nonumber 
\end{align}
That is, in this convention the matter field is minimally coupled in the
Einstein frame, as a consequence of which the matter energy-momentum tensor is
conserved in this frame. Similar approach for the Einstein frame can be found
in~\cite{wetterich14a,wetterich13}. The scalar field $\varphi$ now takes the role of a
quintessence field. Note that, once the quintessence model is specified in the
Einstein frame, the Jordan frame action is fixed up to the choice of the
Brans-Dicke parameter $\wbd$. The Brans-Dicke potential is determined from the
quintessence potential as in~\cref{eq:9}, where the fields themselves are
related via~\cref{eq:37}, these relations are the same as in the case of the
concordance model.

Let us consider non-relativistic matter (dust) and radiation components in the
Einstein frame, along with the quintessence field $\varphi$. The above action leads to
the standard Friedmann equation in the Einstein frame
\begin{align}
  \label{eq:45}
  \tilde{H}^2 &= \frac{\kappa^2}{3} \left(  \rho_\varphi(\at) + \rho_M(\at) + \rho_R(\at)\right), 
\end{align}
where $\rho_\varphi$, $\rho_M$, and $\rho_R$ are the energy densities corresponding to the
quintessence field, non-relativistic matter and radiation, respectively.
Starting with~\cref{eq:19} for Brans-Dicke theory (\cref{eq:32}) and using the
above Friedmann equation we find
\begin{subequations}
  \label{eq:46}
  \begin{alignat}{3}
    \label{eq:47}
    \diff{a}{\at} < 0 &\iff  1 + w_\varphi(\at) &&> \mathcal{C}(\at; \varpi) &&\iff \text{ Collapsing Jordan frame }\\
    \label{eq:48}
    \diff{a}{\at} > 0 &\iff  1 + w_\varphi(\at) &&< \mathcal{C}(\at; \varpi) &&\iff \text{ Expanding Jordan frame}\\
    \label{eq:49}
    \diff{a}{\at} = 0 &\iff  1 + w_\varphi(\at) &&= \mathcal{C}(\at; \varpi) &&\iff \text{ Bounce/collapse in the Jordan frame},
  \end{alignat}
\end{subequations}
where
\begin{align}
  \label{eq:50}
  \mathcal{C}(\at; \varpi) = \frac{2}{3} \varpi \left( 1 + \frac{\Omega_M(\at)}{\Omega_\varphi(\at)} + \frac{\Omega_R(\at)}{\Omega_\varphi(\at)} \right),
\end{align}
$\Omega_{M,R}(\at) = \rho_{M,R}/\rho_c$ are the time-dependant density parameters of dust
and radiation in the Einstein frame. We see that whether the Jordan frame with
$\varpi$ is expanding or collapsing is determined by the equation of state of
the quintessence field and energy densities of all the components present in the
Einstein frame. The Jordan frame goes through a `turn-around', i.e., a bounce
or a collapse, when the equation of state satisfies~\cref{eq:49}. From this we
define
\begin{align}
  \label{eq:51}
  \varpi_*(\at_{\text{TA}}) = \frac{3}{2} \left( 1 + w_\varphi(\at_{\text{TA}}) \right)  \left( 1 + \frac{\Omega_M(\at_{\text{TA}})}{\Omega_\varphi(\at_{\text{TA}})} + \frac{\Omega_R(\at_{\text{TA}})}{\Omega_\varphi(\at_{\text{TA}})} \right)^{-1},
\end{align}
such that the Jordan frame with $\varpi = \varpi_* (\at_{\text{TA}})$
corresponding to the quintessence field goes through a turn-around at the
Einstein frame scale factor $\at=\at_\text{TA}$. We see that
$\varpi_*(\at_{\text{TA}})>0$ for all $\at_{\text{TA}}$ (as long as
$w_\varphi(\at_{\text{TA}}) > -1$, we will ignore cases with $w_\varphi < -1$ as they lead
to the phantom regime). Therefore, a positive $\varpi_*(\at_{\text{TA}})$ always
exists for any $\at_{\text{TA}}$, which is required for the quintessence field
to be real (see~\cref{eq:7}). From these we can conclude the followings:
\begin{enumerate}
\item Quintessence models with standard matter components in the Einstein frame
  \emph{always} correspond to a Jordan frame governed by a Brans-Dicke model
  where the Jordan frame goes through a turn-around, i.e., a bounce or a
  collapse.
  
\item The point of the Jordan frame turn-around, i.e., the value of the Einstein
  frame scale factor at the time of the Jordan frame bounce or collapse, can be
  arranged \emph{anywhere}, determined by the choice of the Brans-Dicke theory
  ($\varpi$).
\end{enumerate}
Therefore, the dual bouncing/contracting universe description is a feature
generic to a variety of late-time cosmological models, including the
quintessence models and the concordance model of the universe. It follows from
above that one value of $\varpi$ is somewhat spacial. All observationally
consistent quintessence models have a common attribute that all of them lead to
the same values of the equation of state parameter and density parameter at the
current epoch in the Einstein frame, $w_{\varphi 0}$, $\Omega_{\varphi 0}$.
Putting $\at = 1$ in~\eqref{eq:51} we define
\begin{align}
  \label{eq:52}
  \varpi_0 = \varpi_*(\at_{\text{TA}} = 1) = \frac{3}{2} \left( 1 + w_{\varphi 0} \right)  \left( 1 + \frac{\Omega_{M 0}}{\Omega_{\varphi 0}} + \frac{\Omega_{R 0}}{\Omega_{\varphi 0}} \right)^{-1}.
\end{align}
For any viable quintessence model, its corresponding Jordan frame with
$\varpi=\varpi_0$ either goes through a bounce or a collapse at the current
epoch ($\at =1$).

\subsection{Turn-around in the Jordan frame: Bounce or collapse?}
\label{sec:bounce-or-crunch}
Whether the Jordan frame turn-around is a bounce or a collapse is determined by
the conditions~\cref{eq:47,eq:48}. At the point of the turn-around, when
$1+w_\varphi(\at) = \mathcal{C}(\at;\varpi)$ is satisfied, if $w_\varphi(\at)$ is decreasing
(increasing) with respect to $\at$, then the Jordan frame is collapsing
(expanding) before the turn-around and expanding (collapsing) after the
turn-around, or the Jordan frame goes through a bounce (collapse). Thus, whether
a quintessence model maps to bouncing Jordan frame or a collapsing Jordan frame
is determined by whether the equation of state of the quintessence field is
decreasing or increasing function of the scale factor at the point of the Jordan
frame turn-around.

Quintessence models are categorized based on the nature of the evolution of the
field, into the \emph{freezing} and \emph{thawing}
types~\cite{amendola.tsujikawa10}. In the \emph{freezing} quintessence models,
the field evolution slows down as time increases and it eventually `freezes' in
the late times, corresponding to a decreasing equation of state parameter which
approaches $w_\varphi \to -1$ in the late times~\cite{chiba.felice.ea13, tsujikawa13,
  zlatev.wang.ea99, amendola.tsujikawa10}. In the \emph{thawing} quintessence
models, the field stays frozen in the early times due to the Hubble friction,
corresponding to $w_\varphi \to -1$. In the late times, the field overcomes the Hubble
friction and starts to evolve, resulting in increasing $w_\varphi$ later on.
Decreasing and increasing equations of state $w_\varphi$ are features generally
associated with freezing and thawing quintessence models, respectively.
Therefore, taking into account the previous arguments, we expect freezing and
thawing quintessence models to be dual to Brans-Dicke Jordan frames with bounce
and collapse, respectively. In the following section we demonstrate these
dualities for viable quintessence models.

\subsection{Examples: Freezing and thawing quintessence models dual to Jordan
  frames with bounce and collapse}
\label{sec:exampl-freez-thaw}
Here we consider examples of freezing and thawing quintessence models and
explicitly show the bouncing and collapsing behaviours of their corresponding
Jordan frames. For both cases, the Jordan frame turn-arounds are arranged at the
current era in the Einstein frame ($\at = 1$). The condition for Jordan frame
turn-around at the current epoch is very weakly dependent on the radiation
contribution, i.e., $\Omega_{R 0}/\Omega_{\varphi 0} \sim 10^{-5}$ is negligible with respect to
$\Omega_{M0}/\Omega_{\varphi 0}$ in~\cref{eq:46}. Therefore, we can safely ignore radiation and
only consider the contribution from non-relativistic matter in the following
examples.

\begin{figure}
  \centering
  \begin{subfigure}{.44\linewidth}
    \includegraphics[width=\textwidth]{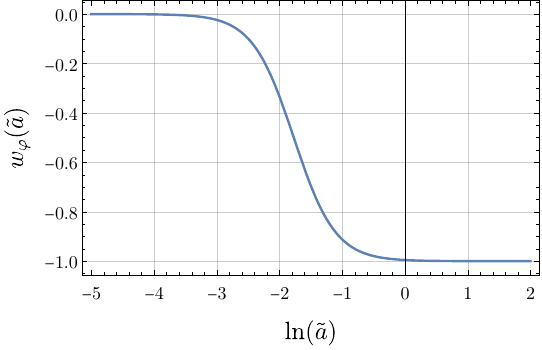}
    \caption{Equation of state parameter.}
    \label{fig:5}
  \end{subfigure}
  \begin{subfigure}{.55\linewidth}
    \includegraphics[width=\textwidth]{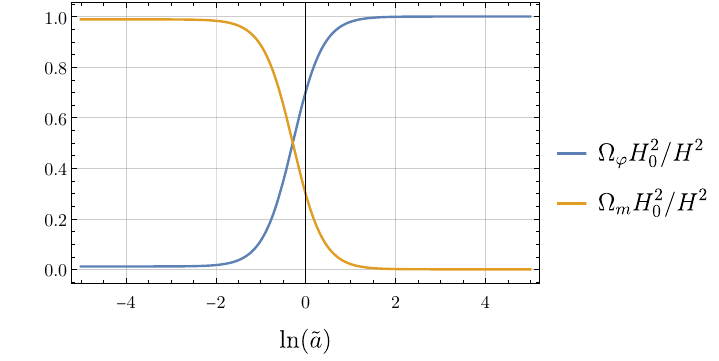}
    \caption{Energy contributions.}
    \label{fig:6}
  \end{subfigure}
  \caption{Equation of state $w_\varphi$ and energy contributions of the quintessence field and
    matter for the freezing quintessence model~\cref{eq:54}, with $w_p=0$, $w_f=-1$, $\at_T = 0.17$,
    $\tau=0.33$. Initially $w_\varphi(\at) \sim 0 $; in the late times as dark energy contribution
    takes over matter contribution, $w_\varphi$ approaches $-1$ asymptotically.}
  \label{fig:7}
\end{figure}

\begin{figure}
  \centering
  \begin{subfigure}{.55\linewidth}
    \includegraphics[width=\textwidth]{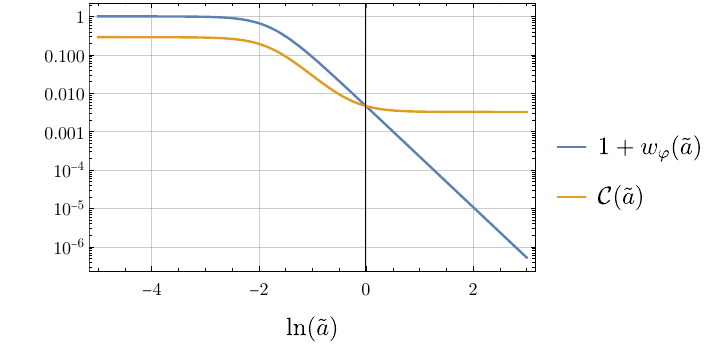}
    \caption{Condition for expanding and collapsing Jordan frame.}
    \label{fig:8}
  \end{subfigure}
  \begin{subfigure}{.44\linewidth}
    \includegraphics[width=\textwidth]{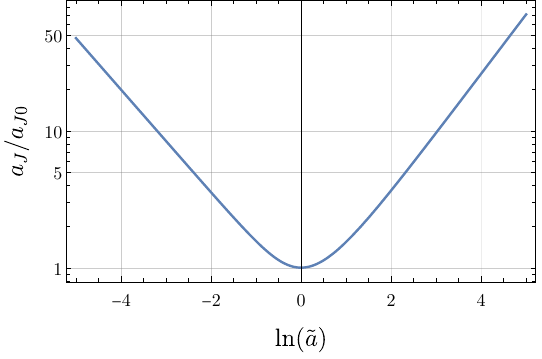}
    \caption{Jordan frame bounce.}
    \label{fig:9}
  \end{subfigure}
  \caption{Bouncing behaviour of the Jordan frame scale factor for freezing quintessence
    model~\eqref{eq:54}, with $w_p=0$, $w_f=-1$, $\at_T = 0.17$; the bounce is arranged at the
    current epoch $\at=1$. (a) Before the current era $1+w_\varphi > \mathcal{C}$ (collapsing Jordan
    frame) and $1 + w_\varphi < \mathcal{C}$ (expanding Jordan frame) after the current era, leading
    to the Jordan frame bounce at the current era. (b) Bouncing behaviour of the Jordan frame scale
    factor $a$, $a_0 = a(\at=1)$.}
  \label{fig:10}
\end{figure}

As an example of the freezing quintessence models, we consider the `double
exponential' potential of the form~\cite{amendola.tsujikawa10,
  chiba.felice.ea13}
\begin{align}
  \label{eq:53}
  V(\varphi) = V_1 \exp \left( - \lambda_1 \kappa \varphi \right) + V_2 \exp \left( - \lambda_2 \kappa \varphi \right).
\end{align}
with the parameters $\lambda_1 \gg 1$ and $\lambda_2 \lesssim 2$. The model leads to an initial
scaling matter era when dark energy is subdominant ($\Omega_\varphi = 3/\lambda_1^2$), followed by
an accelerating dark energy dominated era.  The equation of state for this model can be approximated
using the following parameterization~\cite{chiba.felice.ea13}
\begin{align}
  \label{eq:54}
  w_\varphi(\at) = w_f + \frac{w_p - w_f}{1 + \left(\frac{\at}{\at_T} \right)^{1/\tau}},
\end{align}
where the parameters $w_p$ and $w_f$ represent the initial and final values of
$w_\varphi$; $\at_T$ and $\tau$ determine the transition from matter era to dark energy
era in the Einstein frame. The evolution of the equation of state parameter and
energy contribution of the field are shown in~\cref{fig:7} for $w_p=0$,
$w_f=-1$. We see that $w_\varphi$ initially stays at $0$ in the matter dominated era;
it decreases later on and eventually approaches $-1$ in the late dark energy
dominated era. In~\cref{fig:8}, we plot the condition for expanding and
collapsing Jordan frame and~\cref{fig:9} shows the bouncing behaviour of the
Jordan frame scale factor. In this example, the Brans-Dicke parameter $\varpi$
is chosen such that the Jordan frame bounce occurs at the current era, i.e.
at $\at=1$. The plots show that before the current era $w_\varphi$ satisfies the
condition for collapse,~\cref{eq:47}, while it satisfies the condition for
expansion,~\cref{eq:48} after the current era, thus leading to a bounce in the
Jordan frame at the current era.


\begin{figure}
  \centering
  \begin{subfigure}{.44\linewidth}
    \includegraphics[width=\textwidth]{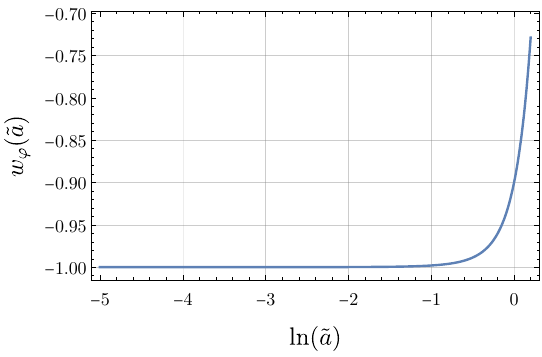}
    \caption{Equations of state parameter.}
    \label{fig:11}
  \end{subfigure}
  \begin{subfigure}{.55\linewidth}
    \includegraphics[width=\textwidth]{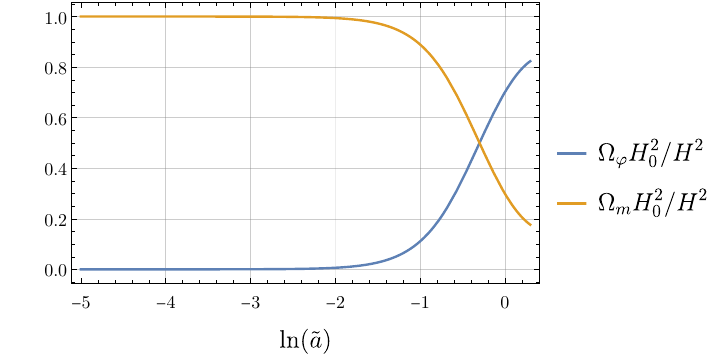}
    \caption{Energy contributions.}
    \label{fig:12}
  \end{subfigure}
  \caption{Equation of state $w_\varphi$ and energy contributions of the quintessence field and
    matter for the thawing quintessence model~\eqref{eq:56}, with $K=2.88$,
    $\Omega_{\varphi 0} = .7$, $w_0=-0.9$. Initially $w_\varphi(\at) \sim 0$; as the dark energy
    takes over in the late times, $w_\varphi$ deviates from $-1$.}
  \label{fig:13}
\end{figure}

\begin{figure}
  \centering
  \begin{subfigure}{.55\linewidth}
    \includegraphics[width=\textwidth]{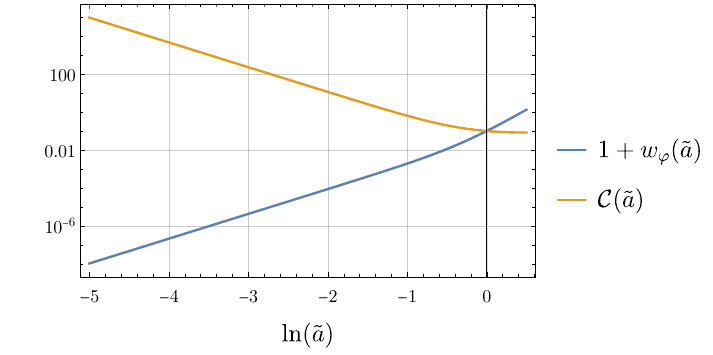}
    \label{fig:14}
  \end{subfigure}
  \begin{subfigure}{.44\linewidth}
    \includegraphics[width=\textwidth]{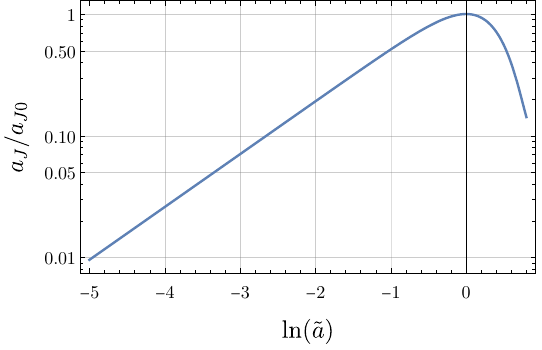}
    \label{fig:15}
  \end{subfigure}
  \caption{Collapsing behaviour of the Jordan frame scale factor for
    the thawing quintessence model~\eqref{eq:56}, with $K=2.88$,
    $\Omega_{\varphi 0} = .7$, $w_0=-0.9$; the collapse is arranged at
    the current epoch $\at=1$. (a) $1+w_\varphi < \mathcal{C}$
    (expanding Jordan frame) before the current era and
    $1 + w_\varphi > \mathcal{C}$ (collapsing Jordan frame) after the
    current era, leading to the Jordan frame collapse at the current
    era. (b) Collapsing behaviour of the Jordan frame scale factor $a$,
    $a_0 = a(\at=1)$.}
  \label{fig:16}
\end{figure}

We now move on to the case of thawing quintessence models. As an example, we
consider the `hilltop' quintessence potential~\cite{chiba.felice.ea13,
  amendola.tsujikawa10}
\begin{align}
  \label{eq:55}
  V(\varphi) = \Lambda^4 \left( 1 + \cos \left( \frac{\varphi}{f} \right) \right).
\end{align}
The field evolution for this potential can approximately be implemented with the
following parameterization of the equation of state~\cite{chiba.felice.ea13}
\begin{align}
  \label{eq:56}
  w_\varphi(\at) = -1 + (1 + w_0) \at^{3(K-1)} \left[\frac{\left(K - F\left(\at\right)\right)\left(F\left(\at\right) + 1\right)^K + \left(K + F\left(\at\right)\right)\left(F\left(\at\right) - 1\right)^K }{\left(K - \Omega_{\varphi 0}^{-1/2}\right)\left(\Omega_{\varphi 0 }^{-1/2} + 1\right)^K + \left(K - \Omega_{\varphi 0}^{-1/2}\right)\left(\Omega_{\varphi 0 }^{-1/2} + 1\right)^K}\right],
\end{align}
where
\begin{align}
  \label{eq:57}
  F(\at) = \sqrt{1 + \left( \Omega_{\varphi 0}^{-1} - 1 \right) \at^{-3}},
\end{align}
$w_0 = w_\varphi(\at=1)$, and $K$ is a constant parameter of the
model. The evolution of $w_\varphi$ and energy contributions from the
quintessence field and matter are shown in~\cref{fig:11,fig:12}, for
$w_0=-0.9$, $\Omega_{\varphi 0}=0.7$, and
$K=2.88$~\cite{chiba.felice.ea13}. The equation of state remains close
to $-1$ in the early times, corresponding to an almost frozen field
$\varphi$. As dark energy contribution takes over the matter
contribution near the current epoch, the field starts to evolve and
$w_\varphi$ increases slightly. The conditions for expanding and
collapsing Jordan frame are plotted in~\cref{fig:11}. We choose the
Brans-Dicke parameter $\varpi$ for the Jordan frame collapse to occur
at the current era. The plot shows that before the current era,
$w_\varphi$ satisfies the condition for expansion,~\cref{eq:48}, while
it satisfies the condition for collapse,~\cref{eq:47} after the
current era, thus leading to a collapse in the Jordan frame at the
current era (\cref{fig:16}).

The duality between an indefinitely contracting Jordan frame and the physical
universe in the Einstein frame has interesting consequence on the conformal
correspondence at the quantum level, similar to the expansion-bounce scenario
discussed in~\cref{sec:jordan-frame:-brans}. Once the scale factor of the
contracting Jordan frame becomes sufficiently small, the Jordan frame universe
is expected to develop non-negligible quantum characteristics. As the Jordan
frame contracts further, the quantum effects therein are expected to grow. While
at the same time, the Einstein frame keeps on expanding with its scale factor
becoming arbitrarily large. It is worth exploring whether the quantum effects in
the Einstein frame are suppressed due to its large scale, or whether the quantum
characteristics are a frame independent effect -- both the conformally connected
universes develop increasing quantum features regardless the cosmological
expansions therein.

\section{De Sitter expansion in the Jordan frame: Collapsing Einstein frame}
\label{sec:de-sitter-expansion}

In the previous sections, we have explored examples of the expansion-collapse
duality, where Einstein frames undergoing accelerating expansion correspond to
bouncing or collapsing Jordan frames. It is, however, also possible to construct
an opposite scenario, where an expanding universe in the Jordan frame can be
described through a collapsing universe in the Einstein frame. In this section,
we briefly discuss an example where a de Sitter expansion in the Jordan frame
maps to a contracting Einstein frame.

Let us consider the prototype Brans-Dicke theory in the Jordan frame, given by
the action in~\cref{eq:34} with $U(\lambda) = 0$. One can show that for the
Brans-Dicke parameter $\wbd = - \frac{4}{3}$, the Jordan frame equations of
motion (see~\cref{a1}) admit de Sitter solution \cite{faraoni04},
\begin{subequations}
  \label{eq:58}
  \begin{align}
  \label{eq:59}
    a(t) &= a_0 \exp \left( H_0 t \right),\\
    \lambda(t) &= \lambda_0 \exp \left( - 3 H_0 t \right),
  \end{align}
\end{subequations}
where $t$ is the Jordan frame coordinate time, $H_0$ is the constant Hubble
parameter in the Jordan frame, $a_0 = a(t=0)$, and $\lambda_0 = \lambda(t=0)$.
In this case, the Jordan frame universe mimics the late-time dark
energy-dominated era of the $\Lambda$CDM model, where, the effect of dark energy
is produced by the Brans-Dicke field instead of the cosmological constant.

In the corresponding Einstein frame (\cref{eq:8}), the potential becomes
$V(\phi) = 0$ (from \cref{eq:9}) and the Einstein frame scalar field describes a
stiff fluid with equation of state parameter $w_\varphi = 1$
(from~\cref{eq:17}). With $\wbd = -4/3$ and $w_\varphi = 1$, the
expansion-collapse duality condition in~\cref{eq:35} is always satisfied.
Therefore, as the Jordan frame expands exponentially, the corresponding Einstein
frame contracts. To see this explicitly, one can reconstruct the solution for
the Einstein frame scale factor through its Jordan frame counterpart.
Using~\cref{eq:58} in~\cref{eq:12}, we find the Einstein frame scale factor as
\begin{align}
  \label{eq:60}
  \tilde{a} = \sqrt{\frac{\kappa^2}{8 \pi} \lambda_0 a_0^3} \left( 1 - \sqrt{\frac{8 \pi}{\kappa^2 \lambda_0}} \frac{3}{2} H_0 \tilde{t} \right)^{\frac{1}{3}} ,
\end{align}
where the Einstein frame coordinate time $\tilde{t}$ is related to the Jordan
frame coordinate time $t$ as (see~\cref{eq:14})
\begin{align}
  \tilde{t} = \frac{2}{3 H_0} \sqrt{\frac{\kappa^2 \lambda_0}{8 \pi}} \left[1 - \exp \left( - \frac{3}{2} H_0 t \right)  \right].
\end{align}
Note that we have chosen the origin of $\tilde{t}(t)$ such that
$\tilde{t}(t = 0) = 0$. In the late-time limit $(t \to \infty)$, the Jordan frame scale
factor $a \to \infty$, while in this, limit the Einstein frame coordinate time
$\tilde{t} \to \frac{2}{3 H_0} \sqrt{\frac{\kappa^2 \lambda_0}{8 \pi}} $ and the scale factor
$\at \to 0$. Therefore, the de Sitter expansion in the Jordan frame can
alternatively be seen in the Einstein frame as a collapsing universe, heading
towards a big crunch-like singularity.

The duality between a de Sitter spacetime and a contracting universe can be a
helpful tool in further studies. For example, quantum fluctuations in the de
Sitter background have been studied extensively as the generator of the
large-scale structure in the universe~(see for
example~\cite{brandenberger1985}). It is well-known that quantum fluctuations of
a massless scalar field in a de Sitter cosmology show divergent behavior in the
infrared limit~(see~\cite{markkanen2018,lochan2018,miao2010} and references
therein). This peculiarity is often associated with the non-existence of de
Sitter invariant vacuum state. It has also been shown that similar divergences
can arise in a universe with power-law expansion~\cite{lochan2018}. One may
address the issue of the diverging quantum fluctuations in the de Sitter and
power-law type expanding spacetimes by posing the problem in a conformally
connected frame that is contracting. When the scale factor in the contracting
universe becomes sufficiently small and approaches the singularity, the
classical description of the system is expected to break down; one may speculate
the universe to develop significant quantum characteristics in this regime.
Therefore, the conformal frame with the contracting universe may provide a
natural framework to study the quantum fluctuation, which, in turn, can be
imported to the conformally connected expanding de Sitter spacetime. The study
of quantum fluctuations in the de Sitter universe through the contracting
universe may provide new insights into the infrared divergence issue, which may
not be apparent otherwise.

\section{Einstein frame-Jordan frame correspondence: Effects of linear perturbations}
\label{sec:einst-jord-frame}

The Einstein and Jordan frame universes are connected via conformal
transformation of the metric. We have so far considered that both the Einstein
and Jordan frame metrics take the form of the spatially flat FRW spacetime, as
given in~\cref{eq:8}. However, in general, both the conformally connected
universes can possess small perturbations on the background of FRW spacetime. A
robust conformal correspondence should provide a regular map between the
perturbations in the Einstein and Jordan frames as well.

The study of cosmological perturbations in several classes of modified theories
of gravity, particularly exploring the Einstein frame-Jordan frame mapping, can
be found in~\cite{hwang90, hwang90-1, hwang97, mukhanov92, hwang.noh96, hwang96,
  hwang.noh01}. In~\cite{hwang90, hwang90-1, hwang97} perturbations in
generalized $f(\phi,R)$ theories are studied using the Einstein frame description.
Cosmological perturbations in the early universe $f(R)$ models is explored, for
example, in~\cite{mukhanov92}.

In this section we introduce linear scalar perturbations in the background
metrics of both the conformally connected frames. Following the treatment
in~\cite{hwang90, hwang90-1, hwang97, mukhanov92}, we briefly review the
relation between metric perturbations in the Einstein and Jordan frames.

\subsection{Metric potentials in Einstein and Jordan frames}

Let us consider scalar perturbations in the Jordan and Einstein frame
metrics. The line elements in the Jordan and Einstein frames, written
in the Newtonian gauge, are
\begin{subequations}
  \label{eq:61}
  \begin{align}
    \d s^2 &= a^2(\eta) \left[ - (1 + 2 \Phi) \d \eta^2 + (1 - 2 \Psi) \delta_{\alpha \beta} \d x^\alpha \d x^\beta \right],\\
    \label{eq:62}
    \d \tilde{s}^2 &= \at^2(\eta) \left[ - (1 + 2 \tilde{\Phi}) \d \eta^2 + (1 - 2 \tilde{\Psi}) \delta_{\alpha \beta} \d x^\alpha \d x^\beta \right],
  \end{align}
\end{subequations}
where $(\Phi, \Psi)$ and $(\tilde{\Phi}, \tilde{\Psi})$ are the metric potentials in the
Jordan and Einstein frames, respectively. Note that we will be using the
conformal time ($\eta$) in this section, which is same for both the
frames.\footnote{As it can be seen from~\cref{eq:12},
  $\d \eta = (1/\at (\te))\d \te= (1/a(t)) \d t$.} The above mentioned perturbed
line elements are related via the conformal transformation~\eqref{eq:3} as
\begin{align}
  \label{eq:63}
  \d \tilde{s}^2 = \Omega^2 \d s^2,
\end{align}
where the conformal parameter is perturbed as well
\begin{align}
  \label{eq:64}
  \Omega(\eta, x^\alpha) = \bar{\Omega}(\eta) + \delta \Omega(\eta, x^\alpha).
\end{align}
Comparing the first order terms in~\cref{eq:63}, one can relate the
metric potentials in the two frames as~\cite{hwang90, hwang90-1,
  hwang97, mukhanov92}
\begin{subequations}
  \label{eq:65}
  \begin{align}
    \label{eq:66}
    \Phi &= \tilde{\Phi} - \frac{\delta \Omega}{ \bar{\Omega}},\\
    \label{eq:67}
    \Psi &= \tilde{\Psi} + \frac{\delta \Omega}{ \bar{\Omega}}.
  \end{align}
\end{subequations}
Let us now come to the example where the Einstein and Jordan frames
are governed by the concordance scalar field ($\varphi$) and the
Brans-Dicke field ($\lambda$), where the fields are now perturbed,
\begin{subequations}
  \label{eq:68}
  \begin{align}
    \lambda \left( \eta, x^\alpha \right) &= \lb(\eta) + \delta \lambda(\eta, x^\alpha),\\
    \label{eq:69}
    \varphi \left( \eta, x^\alpha \right) &= \conb(\eta) + \delta \varphi(\eta, x^\alpha).
  \end{align}
\end{subequations}
Since the Einstein frame is free from anisotropic stress, we further
take $\tilde{\Phi} = \tilde{\Psi}$~\cite{mukhanov92}. The relations
between the metric potentials in the two frames then reduce to
\begin{subequations}
  \label{eq:70}
  \begin{align}
    \Phi &= \tilde{\Phi} - \frac{\delta \lambda}{2 \bar{\lambda}},\\
    \Psi &= \tilde{\Phi} + \frac{\delta \lambda}{2 \bar{\lambda}},
  \end{align}
\end{subequations}
whereas, the two potentials in the Jordan frame are related by
\begin{align}
  \label{eq:71}
  \Psi = \Phi + \frac{\delta \lambda}{\lb}.
\end{align}
This shows that unlike the Einstein frame, the two metric potentials in the
Jordan frame are not equal, which is a well-known characteristic of modified
theories of gravity~\cite{hwang90, mukhanov92}.

We now replace the Brans-Dicke field terms ($\lb, \delta \lambda$) in~\cref{eq:70} with
those of the concordance field. Noting that $\varphi$ and $\lambda$ are related as
(see~\cref{eq:7})
\begin{align}
  \label{eq:72}
  \diff{\varphi}{\lambda} = \sqrt{K[\lambda]} = \frac{1}{\sqrt{2 \kappa^2}} \frac{\sqrt{\varpi}}{\lambda},
\end{align}
we get
\begin{subequations}
  \label{eq:73}
  \begin{align}
    \frac{\delta\lambda}{\lb}  &= \frac{\sqrt{2}\kappa}{\sqrt{\varpi}} \delta \varphi,\\
    \label{eq:74}
                               &=\frac{\sqrt{2}}{\sqrt{\varpi}} \frac{2}{\kappa \conb'} \left( \tilde{\Phi}' + \tilde{\cH} \tilde{\Phi} \right),
  \end{align}
\end{subequations}
where the prime denotes derivative with respect to the conformal time $\eta$,
$\tilde{\cH} =\at'/\at$. The last line is derived using the space-time component
of the perturbed Einstein equation in the Einstein frame~\cite{mukhanov92},
\begin{align}
  \label{eq:75}
  \tilde{\Phi}' + \tilde{\cH}\tilde{\Phi} = \frac{\kappa^2}{2}\conb' \delta \varphi.
\end{align}
Using~\cref{eq:74} in~\cref{eq:70} we can write
\begin{subequations}
  \label{eq:76}
  \begin{align}
    \Phi &= \tilde{\Phi} - \frac{\sqrt{2}}{\sqrt{\varpi}} \frac{1}{\kappa \conb'} \left( \tilde{\Phi}' + \tilde{\cH} \tilde{\Phi} \right)\\
    \Psi &= \tilde{\Phi} + \frac{\sqrt{2}}{\sqrt{\varpi}} \frac{1}{\kappa \conb'} \left( \tilde{\Phi}' + \tilde{\cH} \tilde{\Phi} \right).
  \end{align}
\end{subequations}
These are the relations between the metric potentials in the two conformal
frames.\footnote{See~\cite{mukhanov92} for similar relations between metric
  potentials in the context of $f(R)$ theories.} Thus, once the evolution of the
metric potential in the Einstein frame is known, along with the background
quantities $\at(\eta)$ and $\conb(\eta)$, one can obtain the solutions for the metric
potentials in the Jordan frame.

It is, however, possible that under certain circumstances the above relations
become singular, breaking the perturbative map between the two conformal frames.
This issue is addressed, for example, in~\cite{paul.ea14, bari.ea19}, in the
context of early universe bouncing $f(R)$ theories. It is shown that the
relation between the metric potentials can diverge when the background scalar
field in the Einstein frame goes through an extremum, i.e., $\conb' \to 0$. The
same issue can potentially arise in the case of Brans-Dicke theory governed
Jordan frame. From the relations~\eqref{eq:76}, we see that if the background
concordance field goes through an extremum at an instant, i.e.\ $\conb' \to 0$,
the Jordan frame potentials $(\Phi, \Psi)$ can diverge, even if the Einstein frame
potential and its derivative ($\tilde{\Phi}, \tilde{\Phi}'$) remain finite. This can
cause the Einstein frame-Jordan frame correspondence to break down in the
perturbative regime. We now check whether such a divergence occurs in the
present model of the bouncing universe dual to the physical universe.

From~\cref{eq:18}, the conformal time derivative of the background field can be written as
\begin{align}
    \label{eq:77}
  \conb' = \at \sqrt{(1 + w_\varphi(\at)) \bar{\rho}_\varphi(\at)}.
\end{align}
For the case of the concordance model in the Einstein frame, this relation
becomes
\begin{align}
    \label{eq:78}
    \conb' =  \sqrt{ \rho_{\text{c}} \frac{1 - \Omega_\Lambda}{\at}},
\end{align}
where we have used~\cref{eq:24,eq:26} to obtain~\cref{eq:78}. This shows that
$\conb'$ is non-zero for any finite value of the Einstein frame scale factor
$\at$. Also from~\cref{eq:30}, $\at$ is finite-valued as long as the coordinate
time $\te$ is finite, hence $\conb'$ never becomes $0$ at an instant. Unlike the
models explored in~\cite{paul.ea14,bari.ea19}, the issue of diverging metric
potentials caused by $\conb' \to 0$ is not present in the current model.

For quintessence models of dark energy, $\fb$ takes the role of the
background quintessence field. The relation in~\cref{eq:77} holds true
for quintessence models as well, given $w_\varphi$,
$\bar{\rho}_ \varphi$ are the equation of state and background energy
density associated with the quintessence field. As long as the
quintessence equation of state remains $w_\varphi(\at) > -1$ at any
finite time, $\fb '$ remains non-zero and the relations~\cref{eq:76}
are non-divergent. For example, for the thawing and freezing types of
quintessence models considered in the last
section,~\cref{eq:54,eq:56}, $w_\varphi \neq -1$ for any finite value
of the scale factor $\at$. For the thawing model, $w_\varphi$
approaches $-1$ in the asymptotic future ($\at \to \infty$); while for
the freezing model, $w_\varphi \to -1$ in the asymptotic past
($\at \to 0$) (see~\cref{fig:5,fig:11}). Therefore, we expect the
Jordan frame-Einstein frame map to behave regularly in the
perturbative regime for quintessence models with $w_\varphi > -1$,
even when the Jordan frame goes through the turn-around.

In the following discussion, we mainly focus on the the example of the
concordance model and explicitly verify the stability of the perturbative
conformal map. To do this, we first numerically solve the Einstein frame metric
perturbation, the result is then transported to the Jordan frame using the
conformal correspondence. For the second case, we solve the perturbations in the
Jordan frame directly, the result is then compared with those obtained via the
Einstein frame. If the conformal map is robust under perturbations, then the
Jordan frame metric potentials obtained in these two cases should be in
agreement.

\subsubsection{Numerical results in the Einstein frame}
\label{sec:numer-results-einst}
Let us first consider the evolution of perturbations in the Einstein frame.
Cosmological perturbations in the presence of a canonical scalar field has been
studied broadly, both in the context of inflationary models and dark energy
models (see, for example,~\cite{mukhanov92, kodama.sasaki84} and references
therein). For the perturbed Einstein frame metric~\eqref{eq:62} and the
perturbed concordance field~\eqref{eq:69}, the Einstein field equation leads
to~\cite{mukhanov92}
\begin{align}
  \label{eq:79}
  \tilde{\Phi}'' + 2 \diff{ }{\eta} \left( \frac{\at}{\conb'} \right) \left( \frac{\at}{\conb'} \right)^{-1} \tilde{\Phi}' - \nabla^2 \tilde{\Phi} + 2 \conb' \diff{ }{\eta} \left( \frac{\tilde{\mathcal{H}}}{\conb'} \right) \tilde{\Phi} = 0,
\end{align}
where $\nabla^2$ is the spatial Laplacian. Following the treatment
of~\cite{mukhanov85, mukhanov92}, the above equation for a Fourier
mode $k$ can be put in a relatively simple form,
\begin{align}
  \label{eq:80}
  u''(\eta) + \left( k^2 - \frac{\theta''(\eta)}{\theta(\eta)}\right) u(\eta) &= 0,
\end{align}
where the first order quantity $u$ and background quantity $\theta$
are defined as
\begin{subequations}
  \label{eq:81}
  \begin{align}
    \label{eq:82}
    u(\eta) &= \at(\eta)\frac{\tilde{\Phi}(\eta)}{\conb'(\eta)},\\
    \theta(\eta) &= \frac{\tilde{\cH}}{\at \conb'}.
  \end{align}
\end{subequations}
For small scale modes, or for large $k$ ($k^2 \gg \theta''/\theta$),
the contribution from the $\theta''/\theta$ term in~\cref{eq:80}
becomes minimal, leading to an oscillatory solution for $u$. The
Einstein frame metric potential is then obtained using
$\tilde{\Phi} = (\conb'/\at) u$, where, from~\cref{eq:78}, the factor
$(\conb'/\at)$ can be shown to decay as $1/\at^{\frac{3}{2}}$. Hence,
for $k^2 \gg \theta''/\theta$, we expect $\tilde{\Phi}$ to have the
profile of a damped oscillator.
\begin{figure}
  \centering
  \includegraphics[width=.5\textwidth]{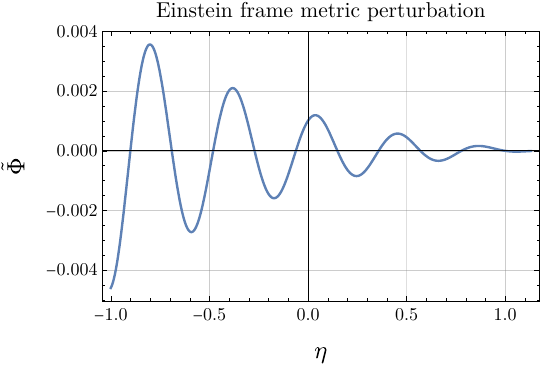}
  \caption{Numerical solution for Einstein frame metric perturbation
    $\tilde{\Phi}$ is plotted with respect to the conformal time $\eta$, for
    Fourier mode $k=15$, where the Einstein frame is governed by the concordance
    scalar field. $\eta$ is given in the unit of $13.5$ Gy, $\eta=0$ corresponds
    to the current Einstein frame epoch, $\at=1$.}
  \label{fig:17}
\end{figure}

The background quantity $\theta(\eta)$ can be obtained analytically, from
\cref{eq:78,eq:30}. Using this, we numerically solve the perturbation
equation~\eqref{eq:80} for such a sufficiently large $k=15$, with initial
conditions
\begin{align}
  \label{eq:83}
  \tilde{\Phi}(\eta = 0) = 10^{-3},\  \tilde{\Phi}'(\eta = 0) = 10^{-2}.
\end{align}
Other background parameters are taken to be
$\Omega_\Lambda=0.7,\ H_0 = 70 \text{ km }\text{sec}^{-1} \text{ Mpc}^{-1}$. The origin of
the conformal time ($\eta = \int \d \te/ \at$) is chosen such that $\eta=0$ coincides
with the current epoch, i.e., $\at(\te=\te_0=0)=\ \at(\eta=0) = 1$ in the Einstein
frame. \Cref{fig:17} shows the numerical evolution of the Einstein frame
perturbation with conformal time. As discussed above, $\tilde{\Phi}$ exhibits
oscillation with decreasing amplitude.

\begin{figure}
  \centering
  \begin{subfigure}{0.49\textwidth}
    \includegraphics[width=\textwidth]{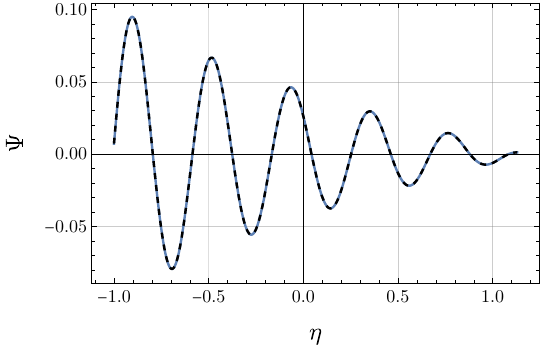}
    \caption{Jordan frame metric potential $\Psi$}
    \label{fig:19}
  \end{subfigure}
  \hfill
  \begin{subfigure}{0.49\textwidth}
    \centering
    \includegraphics[width=\textwidth]{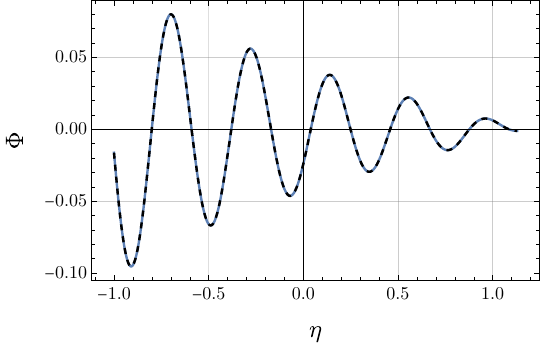}
    \caption{Jordan frame metric potential $\Phi$}
    \label{fig:20}
  \end{subfigure}
  \caption{Jordan frame metric potentials $\Psi$ and $\Phi$ are plotted with
    respect to the conformal time $\eta$ (in the unit of $13.5$ Gy), where
    Einstein frame is governed by the concordance field. Here $\eta=0$ is the
    point of the bounce. In both the figures, the blue plots show numerical
    solutions of the metric potentials obtained directly in the Jordan frame,
    whereas, the black dashed plots are the corresponding results obtained from
    Einstein frame, using the conformal correspondence. The Jordan and Einstein
    frame solutions are in good agreement throughout the evolution. }
  \label{fig:21}
\end{figure}

Once the evolution of $\tilde{\Phi}$ is known, one can obtain the
Jordan frame metric perturbations $\Phi,\ \Psi$, from the
relations~\eqref{eq:76}.  For the numerical results, we choose the
Jordan frame with Brans-Dicke parameter
$\wbd = - \frac{3}{4} \left( 1 + \Omega_\Lambda \right)$. For this
choice of $\wbd$, the Jordan frame bounce occurs exactly at the
current epoch of the Einstein frame, i.e., at $\at=1$. The numerical
evolutions of $\Phi$ and $\Psi$ are plotted in \cref{fig:19,fig:20}.
\begin{figure}
  \centering
  \includegraphics[width=.5\textwidth]{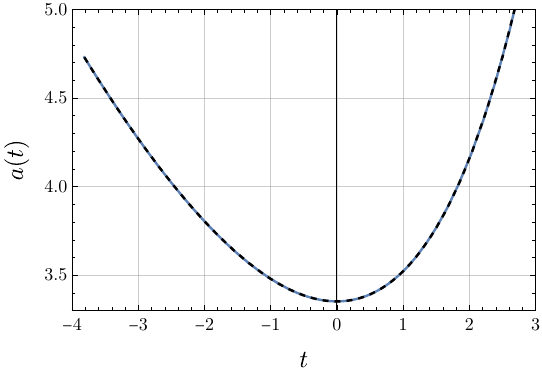}
  \caption{Jordan frame scale factor $a$ is plotted with respect to the Jordan
    frame coordinate time $t$, dual to the concordance model in the Einstein
    frame. Here $\wbd = - \frac{3}{4} \left( 1 + \Omega_\Lambda \right)$, $t$ is
    given in the unit of 13.5 Gy. The Jordan frame bounce occurs at $t=0$. The
    blue plot is the numerical solution obtained directly in the Jordan frame,
    whereas, the black dashed plot is the analytical results obtained via the
    Einstein frame. The Einstein and Jordan frame results are shown to be in
    agreement.}
  \label{fig:18}
\end{figure}
\subsubsection{Numerical results in the Jordan frame}
\label{sec:numer-results-jord}
Having solved the Jordan frame perturbations using the conformal correspondence,
we now study the evolution of perturbations directly in the Jordan frame. We
then compare these results with those obtained via the Einstein frame in the
previous section. If the Einstein frame-Jordan frame correspondence remains
valid in the linear perturbation regime, then the solutions obtained for the two
cases are expected to be in agreement.

Let us first consider the background evolution of the Brans-Dicke
Jordan frame. Starting with the action~\eqref{eq:34} and using the
background metric~\eqref{eq:11}, one can obtain the background
equations of motion for the Brans-Dicke field and the Jordan frame
scale factor as (see~\cref{a1})
\begin{subequations}
  \label{eq:84} 
  \begin{align}
    \label{eq:85}
    \dot{H} &= - \frac{\wbd}{2} \left( \frac{\dot{\lb}}{\lb} \right)^2 + 2 H \frac{\dot{\lb}}{\lb} + \frac{1}{2 \varpi \lb} 16 \pi \left[ \lb U_{,\lambda}(\lb) - 2 U (\lb) \right]\\
    \label{eq:86}
    \ddot{\lb} + 3 H \dot{\lb} &= - \frac{16 \pi}{\varpi} \left[ \lb U_{, \lambda}(\lb) - 2 U(\lb)  \right],
  \end{align}
\end{subequations}
where the overdots represent derivatives with respect to the Jordan frame
coordinate time $t$, $H = \dot{a}/a$ is Jordan frame Hubble parameter. The
Brans-Dicke potential $U(\lambda)$, dual to the concordance potential, is given
in~\cref{eq:39}. Using this, we numerically solve the coupled \cref{eq:85,eq:86}
simultaneously to obtained $a(t)$ and $\lb(t)$. For this and all following
numerical calculations, the origin of the Jordan frame coordinate time ($t$) is
taken such that it coincides with the Einstein frame coordinate time at the
current epoch, i.e., $\te_0=0,\ t(\te=\te_0) = 0,\ \eta(\te=\te_0)=0$. The initial
conditions are fixed it $t=0$ such that they coincide with the Einstein frame
solutions. \Cref{fig:18} shows the evolution of the Jordan frame scale factor
with respect to the Jordan frame coordinate time $t$. For the chosen Brans-Dicke
theory, $\wbd = - \frac{3}{4} \left( 1 + \Omega_\Lambda \right) $, the Jordan frame bounce
occurs at $t=0$, which corresponds to the current epoch in the physical
universe. The same plot also shows the analytical solution for $a(t)$ obtained
using the Einstein frame (from~\cref{eq:41}, using~\cref{eq:14,eq:30}). We see
that the numerical solution obtained directly in the Jordan frame is in
agreement with its counterpart, analytically obtained in the Einstein frame.

Having obtained the background evolution in the Jordan frame, we now
move on to the perturbations. For the perturbed Brans-Dicke field
\begin{align}
  \label{eq:87}
  \lambda = \bar{\lambda}(t) + \delta \lambda(t, x^\alpha),
\end{align}
and the perturbed Jordan frame metric in the Newtonian gauge (written in terms of the coordinate
time $t$),
\begin{align}
  \label{eq:88}
  \d s^2 = -\d t^2 (1 + 2 \Phi)+ a^2(t)  (1 - 2 \Psi) \delta_{\alpha \beta} \d x^\alpha \d x^\beta,
\end{align}
the linear order equations of motions, governing $\delta \lambda$ and $\Phi$, can be written for
Fourier mode $k$ as (see \cref{a1})
\begin{align}
  \label{eq:89}
  \ddot{\delta \lambda }
  +\dot{\delta \lambda } \left(3 H+\frac{\dot{\lb }}{\lb}\right)
  +\delta \lambda  \left(\frac{k^2}{a^2}+\frac{6 H \dot{\lb
  }}{\lb }-\frac{\dot{\lb }^2}{\lb ^2}+\frac{2
  \ddot{\lb }}{\lb }+\frac{16 \pi  \lb 
  U_{,\lambda \lambda}}{\varpi }-\frac{16 \pi  U_{\lambda
  }}{\varpi }\right) \nonumber\\
  -6 H \dot{\lb } \Psi -2 \Psi  \ddot{\lb}-4 \dot{\lb } \dot{\Psi } = 0,
\end{align}
and
\begin{align}
  \label{eq:90} 
  -6 \lb  \varpi  \ddot{\Psi }
  -30 H \lb  \dot{\Psi } \varpi     +\Psi \left(-\frac{2 k^2 \lb  \varpi }{a^2}-24 H^2 \lb  \varpi -12
  \dot{H} \lb  \varpi -\frac{2 \dot{\lb }^2 w \varpi }{\lb
  }\right)
  \nonumber\\
  +\delta \lambda  \left(-\frac{2 k^2 \varpi }{a^2}+36 H^2 \varpi -\frac{6 H
  \dot{\lb } \varpi }{\lb }+18 \dot{H} \varpi -48 \pi  \lb 
  U_{,\lambda \lambda }-48 \pi  U_{,\lambda}-64 \pi  w U_{,\lambda
  }+\frac{\dot{\lb }^2 w \varpi }{\lb ^2}\right) \nonumber \\
  +\dot{\delta \lambda }
  \left(6 H \varpi +\frac{2 \dot{\lb } w \varpi }{\lb }\right)
  = 0
\end{align}
For the Fourier mode $k=15$, the coupled~\cref{eq:89,eq:90} are
numerically solved together to obtain $\Psi(t)$ and
$\delta \lambda(t)$. We choose the following initial conditions at
$t=0$,
\begin{align}
  \label{eq:91}
  \Psi(0)= 0.0263,\  \dot{\Psi}(0)=-0.1623,\ \delta \lambda(0)=0.1130,\ \dot{\delta \lambda}(0)=-0.6745,
\end{align}
such that they coincide with the Einstein frame initial conditions~\eqref{eq:83}
at $\eta = t =0$. Using the solutions of $\Psi(t)$ and $\delta \lambda(t)$, the other Jordan
frame metric potential $\Phi(t)$ can be obtained from~\cref{eq:71}.
\Cref{fig:19,fig:20} show the evolution of the Jordan frame metric perturbations
$(\Phi, \Psi)$, together with their counterparts obtained from the Einstein frame.
Note that the Jordan frame solutions $\Psi(t)$, $\Phi(t)$ are converted to functions
of the conformal time $\eta$, in order to compare them with the Einstein frame
solutions. We see that for both the metric potentials, the Jordan and Einstein
frame solutions are in good agreement.

The Einstein frame solutions behave regularly through out the evolution,
including the point of the Jordan frame bounce at $\eta=0$. The conformal map
between the Einstein and Jordan frames is thus able to accommodate for
cosmological perturbations present in the two frames. Therefore, the bouncing
universe description of the concordance model of cosmology is shown to be stable
against linear perturbations.

\begin{figure}
  
  \centering
  \begin{subfigure}{0.49\textwidth}
    \includegraphics[width=\textwidth]{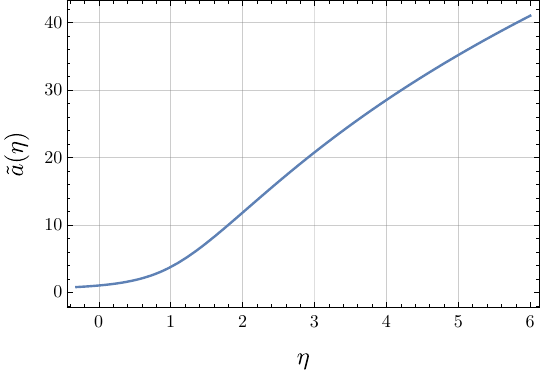}
    \caption{Einstein frame scale factor}
    \label{fig:22}
  \end{subfigure}
  \hfill
  \begin{subfigure}{0.49\textwidth}
    \includegraphics[width=\textwidth]{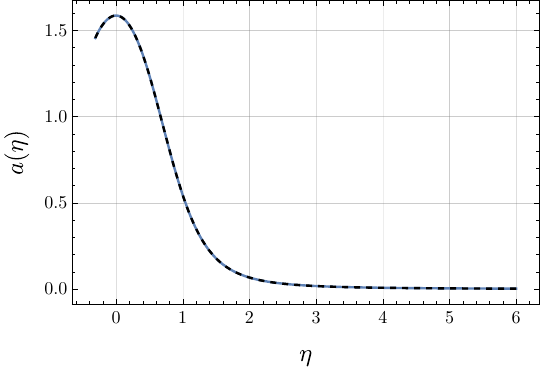}
    \caption{Jordan frame scale factor}
    \label{fig:23}
  \end{subfigure}
  \\ \bigskip
  \begin{subfigure}{0.49\textwidth}
    \includegraphics[width=\textwidth]{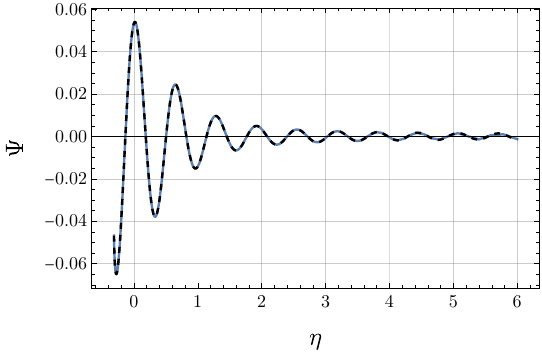}
    \caption{Jordan frame metric potential $\Psi$}
    \label{fig:24}
  \end{subfigure}
  \hfill
  \begin{subfigure}{0.49\textwidth}
    \centering
    \includegraphics[width=\textwidth]{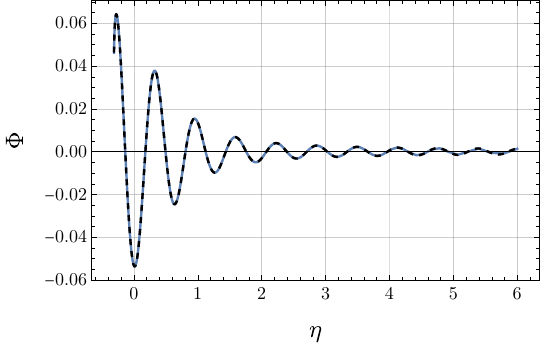}
    \caption{Jordan frame metric potential $\Phi$}
    \label{fig:25}
  \end{subfigure}
  \caption{Conformal duality at the perturbative regime between the Einstein
    frame with a quintessence field (without matter) and its dual collapsing
    Jordan frame with a Brans-Dicke theory. The equation of state of the
    quintessence field is $w_\varphi(\tilde{a}) = w_0 - w' \ln \tilde{a}$, where
    $(w_0, w') = (-0.87, -0.48)$~\cite{sangwan.mukherjee.ea18}. All the plots
    are with respect to the conformal time $\eta$ (in the unit of $13.5$ Gy).
    The Brans-Dicke parameter is chosen such that the Jordan frame turn-around
    occurs at $\eta=0$, $\tilde{a} = 1$. In (a) and (b), the Einstein and Jordan
    frame scale factors ($\tilde{a}$ and $a$) are plotted with respect to the
    conformal time. Jordan frame metric potentials $\Psi$ and $\Phi$ are plotted
    for $k=10$ in (c) and (d). In both the figures, the blue plots show
    numerical solutions of the metric potentials obtained directly in the Jordan
    frame, whereas, the black dashed plots are the corresponding results
    obtained from Einstein frame, using the conformal correspondence. The Jordan
    and Einstein frame solutions are in good agreement throughout the evolution,
    even as $a \to 0$.}
  \label{fig:26}
  
\end{figure}

As the final example, we carry out the same analysis for a collapsing
Jordan frame (see~\cref{fig:26}). For this, we consider a quintessence
field in the Einstein frame with logarithmic equation of state
parameter,
\begin{align}
  \label{eq:104}
  w_\varphi(\tilde{a}) = w_0 - w' \ln \tilde{a},
\end{align}
where $(w_0, w')$ are constant parameters of the
model~\cite{sangwan.mukherjee.ea18,tripathi.sangwan.ea17,efstathiou99}. We
ignore the contribution from the ordinary matter component in this example. For
$w'<0$, the equation of state is an increasing function of the scale factor,
therefore the corresponding Jordan frame undergoes a collapse. After the
collapse, the Jordan frame scale factor continues to contract indefinitely
(\cref{fig:23}), while the Einstein frame expands (\cref{fig:22}). We perform
the same numerical analysis as before and show that the map between the Einstein
and Jordan frame perturbations remains well-behaved at the point of the
turn-around, and even as the Jordan frame scale factor
$a \to 0$~(\cref{fig:24,fig:25}).

\section{Summary and discussion}
\label{sec:discussion}

The scalar-tensor theories belong to a class of modified theories of gravity,
which can be mapped to Einstein's gravity using the Jordan frame-Einstein frame
duality. The Einstein and Jordan frame descriptions are mathematically
equivalent, it is, however, possible that the Jordan frame universe goes through
a quite distinct cosmological evolution in comparison with its Einstein frame
counterpart~\cite{ briscese.elizalde.ea07, paul.ea14, wetterich13, wetterich14a,
  ijjas.ea15, fertig.ea16, graef.ea17, bhattacharya.ea16,
  bahamonde.odintsov.ea16, bahamonde.odintsov.ea17, francfort.ea19, bari.ea19,
  mukherjee.jassal.ea21}. In this paper, we study a class of dual universes with
contrasting nature of evolutions. We consider an Einstein frame where the
universe goes though the standard late-time cosmological evolution, driven by
dark energy and non-relativistic matter. Dual to this, we find a class of
scalar-tensor theories which may lead to a collapsing Jordan frame. A general
condition for such an expansion-collapse duality is obtained to predict whether,
for a given scalar-tensor theory, the Jordan frame is collapsing during a period
of evolution.

We first set up the Einstein frame scalar field for it to provide an effective
description of the $\Lambda$CDM or the concordance model of cosmology. As an example
of the scalar-tensor theories, we consider the Brans-Dicke model in the Jordan
frame. Using the general condition for the expansion-collapse duality, we show
that concordance model of cosmology can always be mapped to a bouncing
Brans-Dicke Jordan frame. This implies that the $\Lambda$CDM model of the late-time
universe can have an effective description in terms of the dynamics of a
bouncing universe. In the second case, the Einstein frame scalar field is
treated as a quintessence field and a non-relativistic matter component is added
separately in the Einstein frame. Quintessence models are shown to be always
dual to Brans-Dicke Jordan frames with turn-arounds, i.e., a bounce or a
collapse. The nature of evolution of the equation of state of the quintessence
field at the turn-around determines whether the turn-around is a bounce or a
collapse. These results indicate that the effect of dark energy can
alternatively be realized as collapse of space in a conformally connected
universe, at least classically.

We have also discussed an example of the expansion-collapse duality where a de
Sitter expansion in the Jordan frame maps to a collapsing Einstein frame heading
towards singularity. The dual description of a de Sitter spacetime through a
collapsing universe has potential applications. For example, one may study the
infrared divergence of quantum fluctuations in the de Sitter
background~\cite{markkanen2018,lochan2018,miao2010} by posing the problem in a
collapsing universe where the quantum effects are better understood.

Previous studies have explored the expansion-collapse duality between conformal
frames in different scenarios~\cite{wetterich13, wetterich14a, ijjas.ea15,
  fertig.ea16, graef.ea17}. For example, in~\cite{wetterich13} (also
see~\cite{wetterich14a}), the Jordan frame universe is contracting in the
radiation and matter dominated phases, while it is expanding in the dark energy
era. In the ``anamorphic'' model, the universe undergoes an initial anamorphic
phase, which exhibits features of both contracting (ekpyrotic) and expanding
(inflationary) models~\cite{ijjas.ea15, graef.ea17}. After the end of the
anamorphic phase the universe enters into the standard hot expanding phase,
consistent with general relativity. In the present work, we explore the
expansion-collapse duality in the context of late-time cosmology. We show that
dual universes with bounce or collapse are not exclusive to specific dark energy
models; in fact, quintessence models, in general, can always be mapped to Jordan
frames with turn-arounds. Moreover, the expanding and contracting phases in the
Jordan frame need not be corresponding to certain eras in the Einstein frame.
The Jordan frame can always be tuned to arrange the turn-around anywhere in the
Einstein frame.

A robust conformal correspondence between the Einstein and Jordan frames should
provide a regular map between cosmological perturbations therein. We investigate
whether the dual bouncing and collapsing universe description of the standard
cosmology is robust against linear perturbations. The conformal correspondence
is found to be stable as long as the Einstein frame scalar field's equation of
state satisfies $w_\varphi>-1$.

To see this explicitly, we primarily consider the example of the concordance
field in the Einstein frame. We numerically solve scalar metric perturbations in
both the conformal frames. The Einstein frame solutions are then transported to
the Jordan frame, using the linear order conformal map. These are compared with
the Jordan frame perturbations, directly obtained in the Jordan frame. The
solutions of the Jordan frame perturbations in these two cases are in agreement.
Therefore, in this case the conformal duality between the late-time physical
universe and the bouncing Brans-Dicke Jordan frame universe is shown to be
stable under linear perturbations. The same analysis is performed for another
example, where the Einstein frame is governed by a quintessence field with
logarithmic equation of state. The corresponding Jordan frame undergoes a
collapse and then contracts indefinitely. The results here also indicate that
the conformal correspondence between the perturbations in the two frames behave
regularly, even as the Jordan frame scale factor $a \to 0$.

Bouncing scenarios, in general, have been explored in the literature as a
candidate for the early universe model, alternative to the inflationary theory.
In this paper we show that a perturbatively stable, effective bouncing
description is also possible for the standard late-time evolution of the
universe. This may have interesting implications. As we have seen, the nature of
the cosmological evolution depends on the conformal frame, however, it is shown
in the literature that physical observables, such as the redshift, galaxy number
count~\cite{francfort.ea19}, Sachs-Wolfe effect, curvature perturbation
(see~\cite{chiba.ea13, prokopec.ea13} and references therein) are independent of
the choice of the conformal frame. One can use these frame invariant observables
in the current model to find a map between the late-time cosmological
perturbations and fluctuations in a bouncing model. Typically, small
perturbations on a collapsing cosmological background tend to grow, whereas the
growth of perturbations is suppressed in an expanding space. Therefore, one can
speculate that the perturbations in the bouncing universe will grow up to the
point of the bounce. On the other hand, the point of the Jordan frame bounce can
be arranged at the onset of the cosmological constant dominated era in the
Einstein frame. The interpretation of the evolution of perturbations in the
bouncing universe may reveal interesting features regarding how dark energy
affects the evolution of cosmological perturbations in the standard cosmology,
which may not be apparent otherwise. Also, since the Jordan frame scale factor
at the point of bounce is sensitive to the dark energy content in the Einstein
frame, such a duality may help in a better understanding of the concordance
model of cosmology.

The status of the conformal correspondence at the quantum level is explored in
the literature in several contexts (see, for example~\cite{ashtekar2003,
  grumiller2002, grumiller2003, fujii1990, flanagan2004, artymowski2013,
  kamenshchik2015, banerjee2016, faraoni.ea07} and references therein). The
duality between the physical universe and bouncing and collapsing universe
presented in this paper may provide further insights into the topic. In the
bouncing and collapsing descriptions of the late-time cosmological models, the
scale factor can become arbitrarily small at a time when the dual universe
represents the accelerating and expanding physical universe. As the collapsing
universe reaches a sufficiently small scale, it is expected to develop quantum
characteristics; for example, the quantum variance in cosmological quantities
may become large in these regions. However, the expanding universe is expected
to behave classically at this point.

It is shown in~\cite{mukherjee2023} that the conformal map between a contracting
Brans-Dicke Jordan frame and an expanding Einstein frame holds at the quantum
level. Interestingly, the rise in quantum fluctuations is found to be
conformally invariant; they take the same value in both frames regardless of the
different cosmological evolutions. Given that the effect of dark energy can
alternatively be realized as collapse of space, as shown in the present work,
the results may indicate that a dark energy-driven large expanding universe can
also harbor non-trivial quantum
features~\cite{dhanuka2020,alexandre2022,dhanuka2022}.


\section*{Acknowledgement}
Research of K.L.\ is partially supported by the DST, Government of India through
the DST INSPIRE Faculty fellowship (04/2016/000571).

\appendix
\section{Field equations in the Jordan frame}
\label{a1}

Here we summarize the derivations of the Brans-Dicke theory equations
of motion in the Jordan frame,~\cref{eq:84,eq:89,eq:90}. For detailed
derivation see, for example~\cite{faraoni04}.

The variation of the Brans-Dicke action~\eqref{eq:34} with respect to
the metric $g^{ab}$ leads to the field equation
\begin{align}
  \label{eq:92}
  \lambda G_{ab} = \frac{\wbd}{\lambda} \left( \nabla_a \lambda \nabla_b \lambda - \frac{1}{2} g_{ab} \nabla^c \lambda \nabla_c \lambda \right) + (\nabla_a \nabla_b \lambda - g_{ab}\Box \lambda) - 8 \pi U g_{ab},
\end{align}
the trace of which is
\begin{align}
  \label{eq:93}
  \lambda R = \frac{\wbd}{\lambda} \nabla^c \lambda \nabla_c \lambda + 3 \Box \lambda + 32 \pi U.
\end{align}
The equation of motion of the Brans-Dicke field $\lambda$ can be derived as
\begin{align}
  \label{eq:94}
  2 \wbd \Box \lambda + \lambda R - \frac{\wbd}{\lambda} \nabla_c \lambda \nabla^c \lambda - 16 \pi \lambda U_{,\lambda} = 0.
\end{align}
Using~\cref{eq:93,eq:94}, one can find the following dynamical
equations for the metric and for $\lambda$,
\begin{subequations}
  \label{eq:95} 
  \begin{align}
    \label{eq:96}
    \varpi  \lambda R &= \varpi  \frac{\wbd}{\lambda} \nabla_c \lambda \nabla^c \lambda + 16 \pi \left( 3 \lambda U_{,\lambda} + 4  \wbd U \right).\\
    \label{eq:97}
    \varpi \Box \lambda &= 16 \pi \left( \lambda U,_\lambda - 2 U \right).
  \end{align}
\end{subequations}
For the homogeneous background field $\lambda(t)$ and spatially flat
FRW line element (written in the coordinate time) in the Jordan frame,
\begin{align}
  \label{eq:98}
  \d s^2 = - \d t^2 + a^2(t) \delta_{\alpha \beta} \d x^\alpha \d x^\beta,
\end{align}
the above equations lead to
\begin{subequations}
  \label{eq:99} 
  \begin{align}
    \label{eq:100}
    \dot{H} &= - \frac{\wbd}{2} \left( \frac{\dot{\lambda}}{\lambda} \right)^2 + 2 H \frac{\dot{\lambda}}{\lambda} + \frac{1}{2 \varpi \lambda} 16 \pi \left[\lambda U_{,\lambda} - 2  U  \right]\\
    \label{eq:101}
    \ddot{\lambda} + 3 H \dot{\lambda} &= -\frac{16 \pi}{\varpi} \left[ \lambda U_{, \lambda} - 2  U  \right].
  \end{align}
\end{subequations}
Similarly, using the perturbed metric~\eqref{eq:88} and perturbed
field~\eqref{eq:87} in~\cref{eq:96,eq:97}, one can find the linear order
equation
\begin{align}
  \label{eq:102}
  -6 \lb  \varpi  \ddot{\Psi }
  -30 H \lb  \dot{\Psi } \varpi     +\Psi \left(-\frac{2 k^2 \lb  \varpi }{a^2}-24 H^2 \lb  \varpi -12
  \dot{H} \lb  \varpi -\frac{2 \dot{\lb }^2 w \varpi }{\lb
  }\right)
  \nonumber\\
  +\delta \lambda  \left(-\frac{2 k^2 \varpi }{a^2}+36 H^2 \varpi -\frac{6 H
  \dot{\lb } \varpi }{\lb }+18 \dot{H} \varpi -48 \pi  \lb 
  U_{,\lambda \lambda}-48 \pi  U_{,\lambda}-64 \pi  w U_{\lambda
  }+\frac{\dot{\lb }^2 w \varpi }{\lb ^2}\right) \nonumber \\
  +\dot{\delta \lambda }
  \left(6 H \varpi +\frac{2 \dot{\lb } w \varpi }{\lb }\right)
  = 0
\end{align}
and
\begin{align}
  \label{eq:103}
  \ddot{\delta \lambda }
  +\dot{\delta \lambda } \left(3 H+\frac{\dot{\lb }}{\lb}\right)
  +\delta \lambda  \left(\frac{k^2}{a^2}+\frac{6 H \dot{\lb
  }}{\lb }-\frac{\dot{\lb }^2}{\lb ^2}+\frac{2
  \ddot{\lb }}{\lb }+\frac{16 \pi  \lb 
  U_{,\lambda \lambda}}{\varpi }-\frac{16 \pi  U_{\lambda
  }}{\varpi }\right) \nonumber\\
  -6 H \dot{\lb } \Psi -2 \Psi  \ddot{\lb}-4 \dot{\lb } \dot{\Psi } = 0,
\end{align}
where all the terms involving $\Phi$ are replaced with $\Psi$ using~\cref{eq:71}.

\end{document}